\documentclass[oldversion]{aa}

\usepackage{amsmath}
\usepackage{natbib}
\usepackage{epsfig}
\usepackage{txfonts}
\usepackage[english]{babel}

\newcommand{\id}{{\,\rm d}}

\newcommand{\beq}{\begin{equation}}   %

\newcommand{\eeq}{\end{equation}}   %

\newcommand{\beqa}{\begin{eqnarray}}   %

\newcommand{\eeqa}{\end{eqnarray}}   %

\newcommand{\beal}{\begin{align}}

\newcommand{\enal}{\end{align}}

\newcommand{\bspl}{\begin{split}}

\newcommand{\espl}{\end{split}}

\newcommand{\bsub}{\begin{subequations}}

\newcommand{\esub}{\end{subequations}}

\newcommand{\bmulti}{\begin{multline}}   %

\newcommand{\beqm}{\begin{mathletters}}   %

\newcommand{\eeqm}{\end{mathletters}}   %

\newcommand{\Abst}[1]{\,#1}

\newcommand{\kB}{k_{\rm B}}

\newcommand{\me}{m_{\rm e}}

\newcommand{\Te}{T_{\rm e}}

\newcommand{\Tg}{T_{\gamma}}

\newcommand{\pd}{\partial}

\newcommand{\pAb}[2]{\frac{\displaystyle\pd #1}{\displaystyle\pd #2}}
\newcommand{\pAbc}[3]{\left.\frac{\displaystyle\pd #1}{\displaystyle\pd #2}\right|_{#3}}

\newcommand{\Abl}[2]{\frac{{\rm d} #1}{{\rm d} #2}}

\newcommand{\pot}[2]{#1 \times 10^{#2}}

\newcommand{\xD}{{{x_{\rm D}}}}
\newcommand{\Erfc}{{\rm Erfc}}
\newcommand{\tauS}{{\tau_{\rm S}}}

\newcommand{\nbb}{{n^{\rm pl}}}

\newcommand{\pdeath}{{p_{\rm d}}}

\newcommand{\psc}{{p_{\rm sc}}}
\newcommand{\nL}{{n_{\rm L}}}

\newcommand{\changeF}[1]{{#1}}
\newcommand{\change}[1]{{#1}}

\begin{document}

\titlerunning{Time-dependent corrections to the Ly $\alpha$ escape probability
  during cosmological recombination}

\title{Time-dependent corrections to the Ly $\alpha$ escape probability
  \\ during cosmological recombination}

\author{J. Chluba\inst{1} \and R.A. Sunyaev\inst{1,2}}
\authorrunning{Chluba \and Sunyaev}

\institute{Max-Planck-Institut f\"ur Astrophysik, Karl-Schwarzschild-Str. 1,
85741 Garching bei M\"unchen, Germany 
\and 
Space Research Institute, Russian Academy of Sciences, Profsoyuznaya 84/32,
117997 Moscow, Russia
}

\offprints{J. Chluba, 
\\ \email{jchluba@mpa-garching.mpg.de}
}

\date{Received 7 October 2008 / Accepted 10 November 2008}

\abstract{We consider the effects connected with the detailed
  radiative transfer during the epoch of cosmological recombination on the
  ionization history of our Universe.
  We focus on the escape of photons from the hydrogen Lyman $\alpha$ resonance
  at redshifts $600\lesssim z \lesssim 2000$, one of two key mechanisms
  defining the rate of cosmological recombination.
We approach this problem within the standard formulation, and corrections due
to two-photon interactions are deferred to another paper.

As a main result we show here that within a non-stationary approach to the
escape problem, the resulting correction in the free electron fraction, $N_{\rm
  e}$, is about $\sim 1.6-1.8\%$ in the redshift range $800\lesssim z \lesssim
1200$. 
Therefore the discussed process results in one of the largest modifications to
the ionization history close to the maximum of Thomson-visibility function at
$z\sim 1100$ considered so far.

We prove our results both numerically and analytically, deriving the escape
probability, and considering both Lyman $\alpha$ {\it line emission} and {\it
  line absorption} in a way different from the Sobolev approximation.
  In particular, we give a detailed derivation of the {\it Sobolev escape
    probability} during hydrogen recombination, and explain the underlying
  assumptions. 
We then discuss the escape of photons for the case of {\it coherent
  scattering} in the lab frame, solving this problem analytically in the {\it
  quasi-stationary approximation} and also in the {\it time-dependent} case.
We show here that during hydrogen recombination the Sobolev approximation for
the escape probability is not valid at the level of $\Delta P/P \sim 5-10\%$.
This is because during recombination the ionization degree changes
significantly over a characteristic time $\Delta z/z\sim 10\%$, so that at
percent level accuracy the photon distribution is not evolving along a
sequence of quasi-stationary stages.
Non-stationary corrections increase the effective escape by $\Delta P/P\sim
+6.4\%$ at $z\sim 1490$, and decrease it by $\Delta P/P\sim -7.6\%$ close to
the maximum of the Thomson-visibility function.
We also demonstrate the crucial role of line emission and absorption in
distant wings (hundreds and thousands of Doppler widths from the resonance) for this effect, and argue that the
final answer probably can only be given within a more rigorous formulation of
the problem using a two- or multi-photon description.
}

\maketitle

\section{Introduction}
\label{sec:Intro}
The extraordinary advances in observations of the Cosmic Microwave Background
(CMB) temperature and polarization angular \changeF{anisotropies}
\citep[e.g.][]{Page2006, Hinshaw2006} and the prospects with the {\sc Planck}
Surveyor\footnote{www.rssd.esa.int/Planck} have motivated several groups to
re-examine the problem of cosmological recombination \citep[e.g. see][for
detailed overview]{Sunyaev2008,Fendt2008}, including subtle physical processes
during hydrogen \citep[e.g. see][]{Dubrovich2005, Chluba2006, Kholu2006,
  Jose2006, Chluba2007b, Hirata2008} and helium recombination
\citep[e.g.  see][]{Switzer2007I, Switzer2007II, Switzer2007III,
  Kholupenko2007, Wong2007, Jose2008, Kholupenko2008}.
It has been argued that percent level corrections to the ionization history
exist, which should be taken into account for future determinations of
cosmological parameters using CMB data obtained with the {\sc Planck}
Surveyor.

In this paper \changeF{we investigate} the validity of one of the key
simplifications used for computations of the hydrogen recombination history
within existing multi-level recombination codes: the {\it
  Sobolev approximation} for the escape of Lyman $\alpha$ photons from the
center of the resonance.
With this approximation it is possible to separate the problem of the
evolution of the photon field and the populations of the hydrogen atom.

Originally the Sobolev approximation was developed in order to describe the
escape of photons from {\it finite} expanding envelopes of planetary nebulae
and stars \citep{Sobolev1960}, but it has been shown that even for
cosmological applications, i.e. {\it infinite} slowly expanding media, it
is very useful \citep{Grachev1991, Hummer1992, RybickiDell94}.
It gives the same answer as less sophisticated methods, based on simple
solutions of the integral or differential equations of radiative transfer,
which were used to solve the cosmological hydrogen recombination problem
in the 1960s \citep{Varshalovich1968, Zeldovich68, Peebles68}.
Both for the Sobolev approximation and these simpler derivations the main
assumptions are:
(i) the properties of the medium (e.g. ionization degree, density, expansion
rate) do not change much while the photons \changeF{interact strongly} with
the Lyman $\alpha$ resonance and (ii) each scattering leads to a {\it
  complete redistribution} of photons over the {\it whole} line profile.

Due to assumption (i) it is possible to approximate the evolution of the
photon distribution as {\it quasi-stationary}, which for conditions in our
Universe seems to be reasonable \citep{RybickiDell94}. However, up to what
level of accuracy remains a difficult question and deserves further
investigations.
On the other hand, assumption (ii) is much less justified, since complete
redistribution requires some process that {\it destroys the coherence} in the
resonance scattering event. This is usually done by {\it collisional
  processes}, which for the conditions in our Universe are extremely
inefficient \citep[e.g. see][]{Chluba2007}.
We will demonstrate here that for present day \changeF{experimental requirements}, i.e. sub-percent level accuracy in the theoretical
predictions of the CMB power spectra at large multipoles $l$ \citep[e.g.
see][]{Seljak2003}, both approximations become insufficient.

\change{In order to understand this problem, it is important that} the
ionization degree during cosmological hydrogen recombination changes with
characteristic time $\Delta z/z\sim 10\%$.
Also, it is clear that photons, which are released in the distant wings of the
Lyman $\alpha$ line\footnote{At redshift $z=1100$ a thousand Doppler width
  corresponds to $\Delta\nu/\nu\sim 2\%$, a distance from the line center that
  can be passed by redshifting in $\Delta z/z \sim 2\%$.}, can in principle
travel, scatter, and redshift for a very long time until being reabsorbed.
Here it is important to distinguish between {\it line scattering} events, and
{\it line emission} and {\it absorption} processes.  
The former only lead to a redistribution of photons over frequency, but {\it
  no} net change in the ionization degree, while the latter (which for example
are connected with direct transitions of electrons between the continuum and
the 2p state) are able to change the number of Lyman $\alpha$ photons, and
hence the ionization degree.
\changeF{Note that} during hydrogen recombination, line absorption
occurs with much lower probability ($\sim 10^{-4}-10^{-3}$) than
line scattering, so that photons only {\it die} or disappear effectively
rather close to the line center (within a few ten to hundred Doppler widths
from the resonance), while in the distant wings they mainly scatter.
\change{
In addition, every photon that was absorbed (or died) will be replaced by a
new photon in a line emission event after a very short time.  The profile of
this line emission is usually described by a Voigt profile, so that the
combination of line absorption followed by a line emission appears to lead to
a complete redistribution of photons over the {\it whole} Lyman $\alpha$ line profile.
However, from the microscopic point of view this is not a scattering event
\footnote{\changeF{\citet{Switzer2007I} also make this distinction using the termini of incoherent processes and coherent scattering.}}.
}

As explained in \citet{RybickiDell94}, in the expanding Universe the
redistribution of photons due to Lyman $\alpha$ resonance scattering is more
accurately described by so-call type-II redistribution \citep{Hummer1962}
rather than by complete redistribution. 
In the former case the photon scatters {\it coherently in the rest-frame of
  the atom}, so that the changes in the energies of the photon after the
scattering event are related to the motion of the atom.
Studying this type of redistribution process in detail is beyond the scope of
this paper, but our computations \citep{Chlubaprep}, in very good agreement
with earlier works \citep[e.g. see][]{RybickiDell94}, show that in a
time-dependent formulation of the problem, including {\it Doppler broadening},
{\it atomic recoil} and {\it stimulated emission}\footnote{Within a
  Fokker-Planck approach the atomic recoil effect was first included by
  \citet{Basko1978a}, while the effect of stimulated emission was only taken
  into account very recently by \citet{Rybicki2006}.}, the actual solution for
the \change{scattered} photon distribution is very close to the one in the
case of {\it no redistribution}, or equivalently {\it no line scattering}.
Here we show in addition that the assumption of complete redistribution leads
to several unphysical conclusions, both in the quasi-stationary approximation
and a time-dependent approach. This is due to the very peculiar conditions in
our Universe, where collisional processes are not important, and in particular
where due to the low Hubble expansion rate, the Sobolev optical depth reaches
extreme values of $\sim 10^6-10^8$ during recombination.

We therefore investigate the evolution of the photon field in the
no-scattering approximation and show that time-dependent corrections to the
effective escape probability are important at the level of $\sim 5\%-10\%$
(see Sect.~\ref{sec:Pesc_def_num_Voigt_time}, Fig.~\ref{fig:DP_P.time.P0} and
Fig.~\ref{fig:DP_P.time}).
As mentioned above, this is due to the fact that in the distant wings of the
Lyman $\alpha$ resonance photons mainly scatter, but do not disappear. This renders
it important to include changes in the ionization degree and photon emission
rate during the evolution of the photon field in the computations, implying
that the quasi-stationary approximation becomes inaccurate.
Both changes in the absorption optical depth and the effective emission rate
cannot be neglected.
The corresponding time-dependent changes in the free electron fraction, which
are important for the Thomson visibility function and in computations of the
CMB power spectra, reach the level of $\sim 1.6-1.8\%$ in the redshift range
$800\lesssim z \lesssim 1200$ (see Sect.~\ref{sec:changes_in_Xe} and
Fig.~\ref{fig:DNe.predict}), and therefore are about 2 times as large as those
due to atomic recoil, recently studied by \citet{Grachev2008}.
Taking the time-dependent correction investigated here into account
will therefore be very important for the analysis of future CMB data from the
{\sc Planck} Surveyor.

\change{ 
We also briefly discuss another aspect of the Sobolev approximation, which is
connected to the {\it shape} of the Lyman $\alpha$ line profile (see
Sect.~\ref{sec:P_shape}).
%
%
In the Sobolev approximation there is no direct dependence of the
Sobolev escape probability on the shape of the line \changeF{emission, absorption, and scattering
profiles}, as long as \changeF{all} are {\it identical}.
Our derivation also clearly shows this point (cf. Sect.~\ref{sec:Sobolev} and
\ref{sec:P_shape}).
Therefore, in principle it does not matter if the profile is assumed to be a
Lorentzian, a Voigt profile, a pure Doppler profile, or a $\delta$ function.
It also turns out that in the no line scattering approximation this is true,
as long as the line emission and absorption profiles are identical, and the
evolution of the photon distribution is quasi-stationary (cf.
Sect.~\ref{sec:ncoh} and \ref{sec:P_shape}).

However, if one includes the deviations from quasi-stationarity, then the
result does depend \changeF{in detail on the Lyman $\alpha$} profile, even
if the line emission and absorption profiles still are the same.
For example, in the case of a pure Doppler profile (very narrow), the problem
of the Lyman $\alpha$ photon escape from the resonance due to the expansion of
the Universe would be practically quasi-stationary, and the
Sobolev approximation should be applicable.
This is because the number of photons emitted and absorbed in the optically
thin region of the Lyman $\alpha$ line is exponentially small, and all the
transfer is happening inside the Doppler core, corresponding to
$\Delta\nu/\nu\sim \text{few}\times 10^{-5}$.

On the other hand in the real problem, Lyman $\alpha$ emission and absorption
also occurs in the distant Lorentz wings (at hundreds and thousands of Doppler
widths) of the resonance.
As we show here, at a percent level the number of these photons is very
important for the value of the effective escape probability (e.g. see
Fig.~\ref{fig:DP_P.part.nS_5}).
This shows that it is crucial to understand the profiles (or
cross-sections) of the considered processes in more detail, and for this
probably a formulation in the two- or multi-photon picture will become
necessary.
Also in principle it should be possible to distinguish between different
redistribution processes for the line scattering event, by measuring the shape
and position of the residual, present day CMB Lyman $\alpha$ distortion.

It is extremely impressive that the standard estimates of the Lyman $\alpha$
escape probability, which were used in the first papers on cosmological
recombination, and the Sobolev approximation give such precise (better
than 5-10\%) answers, \changeF{even though} they are based on two \changeF{incorrect} assumptions as
mentioned above.
It is well known that the principal difference (from a physical point of view)
between the cases of partial and complete redistribution does not influence
the final result very much in the majority of astrophysical applications
\citep{Ivanov1973}. However, the enormous requirements of accuracy of
theoretical estimates in the era of precise cosmology change the situation,
and force us to search for percent level corrections to the escape of
Ly $\alpha$ photons from resonance during the epoch of cosmological
recombination.  }

\section{Transfer equations for the photon field}
\label{app:kin_eq}
In this Section we provide the transfer equation describing the evolution of
the photon distribution in the vicinity of the Lyman $\alpha$ resonance. We
include the effect of {\it line emission} and {\it line absorption} in the
expanding Universe for the cases of {\it coherent line scattering} in the
lab frame, and {\it complete redistribution}.
Here we envision all processes as $1+1$ photon processes, as in the
Seaton-cascade description \citep{Seaton1959}, but leave the treatment of
correction due to two-photon interactions for a future paper.
Also the effects of partial frequency redistribution will be discussed in
separate paper.
In Sect.~\ref{sec:TEq_coh} and \ref{sec:TEq_CR} we give the time-dependent
solutions of these equations.
We will use these results in Sect.~\ref{sec:Ly-a-problem} to deduce the
Lyman $\alpha$ escape probability, which then can be utilized to estimate the
corrections to the cosmological ionization history.

\subsection{General kinetic equation for the photon field}
\label{app:kin_eq_gen}
To follow the evolution of the photon field in the expanding Universe we start
with the kinetic equation for the function $N_{\nu}=I_{\nu}/h\nu$, where
$I_{\nu}$ is the {\it physical} specific intensity of the isotropic, ambient
radiation field \citep[e.g. see][]{RybickiDell94}:
\beal
\label{app:kin_eq_N_nu}
\frac{1}{c}\left[
\left.\pAb{N_{\nu}}{t}\right|_{\nu}+2 H N_{\nu}-H\,\nu \left.\pAb{N_{\nu}}{\nu}\right|_{t}
\right]=\mathcal{C}[N_{\nu}]
\Abst{.} 
\end{align}
Here $H(z)$ is the Hubble parameter as a function of redshift $z$ and
$\mathcal{C}[N_{\nu}]$ is the collision term, which describes the emission,
absorption and frequency redistribution processes.
%


%
In order to simplify the left hand side of the equation
\eqref{app:kin_eq_N_nu} we transform to the frequency variable $x=\nu/(1+z)$,
so that
\beal
\label{app:N_nu_N_x}
N_{\nu}=\Abl{x}{\nu}\,N_{x}=\frac{N_{x}}{(1+z)}
\Abst{.} 
\end{align}
Inserting this into Eq. \eqref{app:kin_eq_N_nu} yields
\beal
\label{app:kin_eq_N_x}
\frac{1}{c}\left[
\left.\pAb{N_{x}}{t}\right|_{\nu}+3 H N_{x}-H\,x\pAbc{N_{x}}{x}{t}
\right]=(1+z)\,\mathcal{C}[N_{\nu}]
\Abst{.} 
\end{align}
To obtain $\left.\id N_{x}/\id t\right|_{\nu}$ one can use the total
differential of $N_{x}$
\beal
\label{app:tot_diff_N_x}
\id N_{x}=\pAbc{N_{x}}{t}{x}\id t+\pAbc{N_{x}}{x}{t}\id x
\end{align}
which with $\left.\id x/\id t\right|_\nu=x\,H$ then gives
\beal
\label{app:N_x_t_const_nu}
\pAbc{N_{x}}{t}{\nu}=\pAbc{N_{x}}{t}{x}+H\,x\pAbc{N_{x}}{x}{t}
\Abst{.} 
\end{align}
Inserting this into Eq. \eqref{app:kin_eq_N_x} one finds
\beal
\label{app:kin_eq_N_x_2}
\frac{1}{c}\left[
\pAbc{N_{x}}{t}{x}+3 H N_{x}\right]=(1+z)\,\mathcal{C}[N_{\nu}]
\Abst{.} 
\end{align}
Here the redshifting term was absorbed due to the choice of the frequency
variable.

The term $3 H N_{x}$ can be eliminated using the substitution
$\tilde{N}_x=N_x/(1+z)^3\equiv N_\nu/(1+z)^2$, so that
Eq. \eqref{app:kin_eq_N_nu} takes the form
\beal
\label{app:kin_eq_Nt_x_2}
\frac{1}{c}\pAbc{\tilde{N}_{x}}{t}{x}
&=\frac{\mathcal{C}[N_{\nu}]}{(1+z)^2}
\Abst{.} 
\end{align}
One can easily verify that in the absence of physical interactions
($\mathcal{C}[N_{\nu}]\equiv 0$), in spite of the Hubble expansion, a
Planckian spectrum is not modified \citep[e.g. see][]{Padmanabhan2002}, so
that it is always possible to directly write
\beal
\label{app:kin_eq_DNt_x_2}
\frac{1}{c}\pAbc{\tilde{N}_{x}}{t}{x}
\equiv
\frac{1}{c}\pAbc{\Delta \tilde{N}_{x}}{t}{x},
\end{align}
where $\Delta \tilde{N}_{x}=\tilde{N}_{x}-\tilde{N}^{\rm Pl}_{x}$ is the
corresponding deviation of the spectrum from a blackbody, which in our
coordinates reads
\beal
\label{app:NPl_x}
\tilde{N}^{\rm pl}_{x}=\frac{2}{c^2}\,\frac{x^2}{e^{h x/k T_0}-1}
\Abst{.} 
\end{align}
Here $T_0=2.725\,$K is the CMB temperature today \citep{Fixsen2002}.

\subsection{line emission and line absorption}
\label{app:em_ab}
Although for conditions in the Universe during cosmological
recombination\footnote{Electron and proton collisions are negligible in
  comparison with radiative processes, like photorecombination or
  photoionization, and bound-bound dipole transitions
  \citep[e.g. see][]{Chluba2008a}.} the resonant scattering optical depth
close to the Lyman $\alpha$ line center exceeds unity by several orders of
magnitude, only {\it real} line emission and absorption lead to a {\it net}
change of the photon number.
If we consider an electron in the ground state of hydrogen which after the
absorption of a photon (say close to the Lyman $\alpha$ resonance) is excited
to the 2p state, then there are two routes out of this level: (i) it can
directly decay back to the ground state, re-emitting a photon with (slightly)
changed frequency, depending on the considered redistribution process, or (ii)
it can be further excited to the {\it continuum} or {\it higher shells}
($n>2$) by the subsequent absorption of a blackbody photon from the CMB.
Only in case (ii) does the number of Lyman $\alpha$ photons really change, while
for (i) the photon simply was scattered.

To describe this aspect of the problem, we use the {\it death probability} or
{\it single scattering albedo}, $\pdeath$, which specifies what fraction of
photons that interact with a hydrogen atom in the 1s state, will really
disappear from the photon distribution.
The scattering probability, $\psc=1-\pdeath$, will then give the fraction of
photons that reappear at a different frequency, and hence only underwent a
scattering rather than a real line absorption.

\subsubsection{Death probability or single scattering albedo}
\label{sec:pdeath}
Including all possible ways in and out of the 2p \changeF{level,} the net change in the
number density of electrons in the 2p level can be written as
\beal
\label{eq:dN2pdt}
\Abl{N_{\rm 2p}}{t}+3H N_{\rm 2p}=\left.\Abl{N_{\rm 2p}}{t}\right|^{\rm Ly-\alpha}_{\rm 2p}+R^{+}_{\rm 2p}-R^{-}_{\rm 2p}N_{\rm 2p},
\end{align}
where $\left.\Abl{N_{\rm 2p}}{t}\right|^{\rm Ly-\alpha}_{\rm 2p}$ denotes the
contribution from the Lyman $\alpha$ transition, which we will specify below
(see Sect.~\ref{sec:Ly-trans}), and
\bsub
\label{eq:R2p}
\beal
R^{-}_{\rm 2p}&=R_{\rm 2p c}+\sum_{i>2\rm p}\frac{g_i}{g_{\rm 2p}} A_{\it i \rm 2p}\,\nbb(\nu_{\it i \rm 2p})
\\
R^{+}_{\rm 2p}&=N_{\rm e}\,N_{\rm p}\,R_{\rm c 2p}+\sum_{i>2\rm p} N_i\,A_{\it i \rm 2p}\,[1+\nbb(\nu_{\it i \rm 2p })].
\end{align}
\esub
Here $N_{\rm e}$ and $N_{\rm p}$ are the electron and proton number densities,
and $\nbb=1/(e^{h\nu/k\Tg}-1)$ is the blackbody photon occupation number,
where $\Tg=T_0\,(1+z)$ is the CMB temperature at redshift $z$.
Furthermore, $A_{\it i \rm 2p}$ denotes the spontaneous dipole transition rate
from level $i$ to the 2p state, $\nu_{\it i \rm 2p }$ the corresponding
transition frequency, and $g_i$ the statistical weight of level $i$.
$R_{\rm c 2p}$ and $R_{\rm 2p c}$ are the photorecombination and
photoionization coefficients of the 2p state, which are computed assuming that
the ambient radiation field is Planckian.
Since except for the Lyman series and the 2s-1s two-photon transition, the
emission of photons during cosmological hydrogen recombination only \changeF{produces}
tiny deviations of the photon distribution from a blackbody spectrum $B_\nu$
\citep[e.g.  see][]{Chluba2006b}, this approximation is possible.
Similarly, we have neglected the spectral distortion in the terms due to
transitions from and to higher levels ($A_{\it i \rm 2p}\,n_\nu\approx A_{\it i
  \rm 2p}\,n^{\rm pl}_\nu$), so that $R^{-}_{\rm 2p}$ becomes completely
independent of the solution for the photon field. However, note that
$R^{+}_{\rm 2p}$ still depends on the solution for the populations, $N_i$, of
the excited levels, and the electron and proton number density.

Omitting electron and proton collisions, the total probability for
Lyman $\alpha$ emission $p_{\rm em}$ is therefore given by
\beal
\label{app:p_em}
p_{\rm em}= \frac{A_{21}}{A_{21}+R^{-}_{\rm 2p}}
\Abst{,} 
\end{align}
where $A_{\rm 21}=\pot{6.27}{8}\,\text{s}^{-1}$ is the spontaneous 2p-1s
transition rate.
The corresponding probability for the {\it death} of photons, i.e. removal of
Lyman $\alpha$ photons or return of 2p electrons to the continuum or higher
levels, is then given by $p_{\rm d}=1-p_{\rm em}$.

Note that in Eq.~\eqref{app:p_em} we directly neglected the effects of
stimulated emission. This approximation is well justified, since close to the
Lyman $\alpha$ transition the photon occupation number
$n_\nu=\frac{c^2\,I_{\nu}}{2h\nu^3}\ll 1$ at all relevant redshifts.

\begin{figure}
\centering 
\includegraphics[width=0.90\columnwidth]
{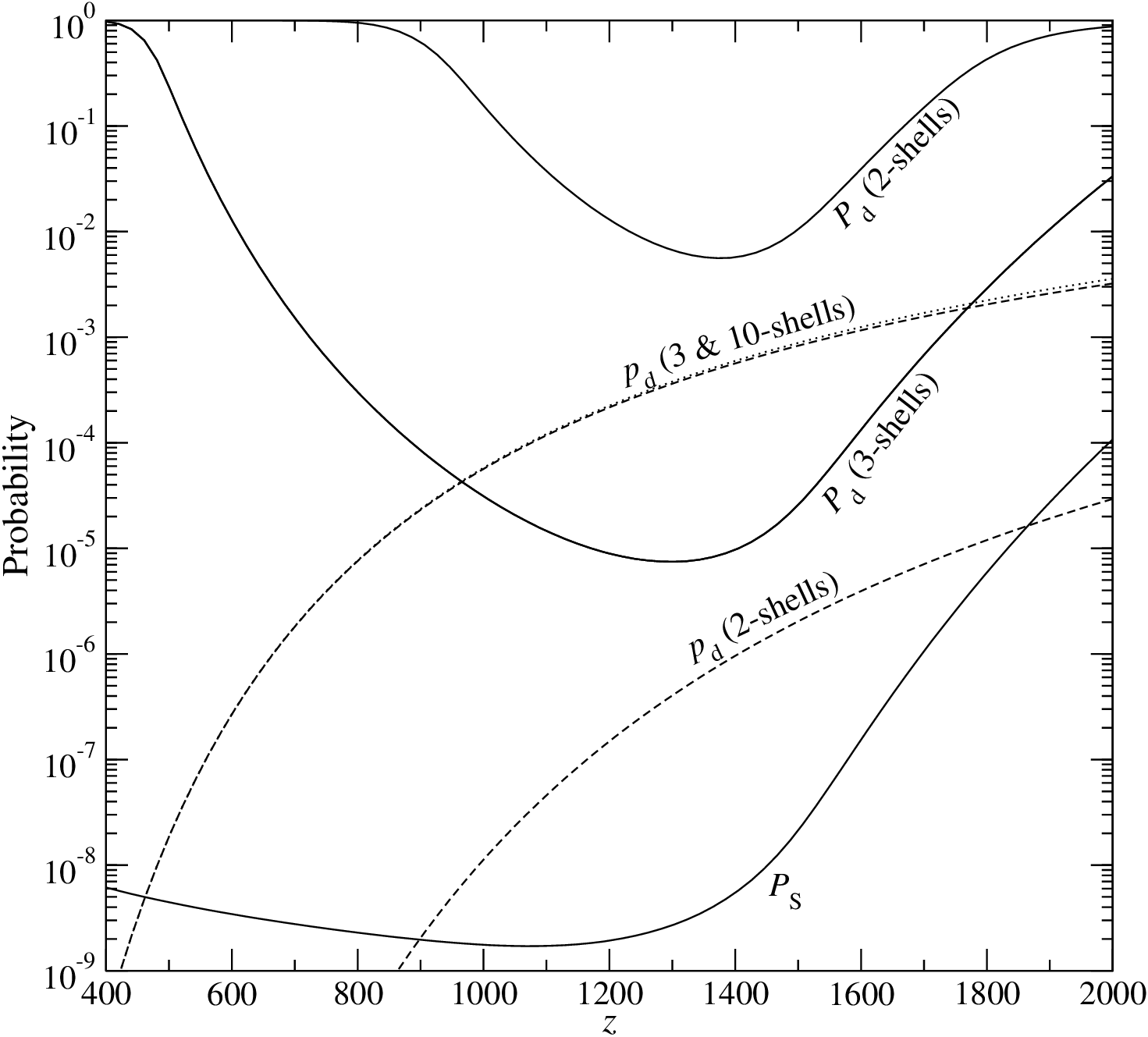}
\caption
{Different probabilities for the cosmological hydrogen recombination problem
  as a function of redshift. The death probabilities, $p_{\rm d}$, for a 2, 3
  (dashed lines), and 10 shell hydrogen atom (dotted line) are shown.  The
  death probability for the 3 shell case already practically coincides with
  the death probability for the 10 shell case.
  The solid lines show different escape probabilities. $P_{\rm
    S}=[1-e^{-\tau_{\rm S}}]/\tau_{\rm S}$ denotes the normal Sobolev escape
  probability, while $P_{\rm d}=[1-e^{-\tau_{\rm d}}]/\tau_{\rm d}$, where
    $\tau_{\rm d}=p_{\rm d}\,\tau_{\rm S}$.
To compute $\tauS$ we used the {\sc Recfast} solution for $N_{\rm 1s}$.  }
\label{fig:p_d}
\end{figure}
In Fig.~\ref{fig:p_d} we show the death probability, $\pdeath$, as a function
of redshift considering a 2, 3 and 10 shell hydrogen atom.
It is clear that the largest contribution to the death probability comes
from the third shell, and cases with $n\geq3$ are practically
indistinguishable.
This is because during cosmological hydrogen recombination $R_{\rm 2p c}\ll
\frac{1}{3}[A_{\rm 3s2p}+5\,A_{\rm 3d2p}]\,\nbb(\nu_{\rm 3s2p})$, and since
$\nbb(\nu_{\rm 3s2p})$, for $n>3$, is exponentially larger than
$\nbb(\nu_{n\rm s2p})$, so that also $[A_{\rm 3s2p}+5\,A_{\rm
    3d2p}]\,\nbb(\nu_{\rm 3s2p})\gg [A_{n\rm s2p}+5\,A_{n\rm
    d2p}]\,\nbb(\nu_{n\rm s2p})$.
This fact implies that for a consistent investigation of the Lyman $\alpha$ escape
\changeF{problem, one should include at least 3 shells} in the computations.

\subsubsection{Line emission profile}
\label{sec:phi_em}
The form of the emission profile for the Lyman $\alpha$ line \changeF{(under the assumption of complete redistribution) is} known
from quantum-mechanical considerations.
Including the thermal motion of the hydrogen atoms it is usually described
using the so-called {\it Voigt} profile:
\bsub
\label{app:Voigt}
\beal
\varphi(\nu)=\frac{a}{\pi^{3/2}\,\Delta\nu_{\rm D}}\int_{-\infty}^\infty
\frac{e^{-t^2}\id t}{a^2+(x_{\rm D}-t)^2}=\frac{\phi(\nu)}{\Delta\nu_{\rm D}}
\Abst{,} 
\end{align}
where for the \ion{H}{i} Lyman $\alpha$ transition the Voigt parameter, $a$,
the Doppler width of the line due to the thermal motion of the hydrogen atoms,
$\Delta\nu_{\rm D}$, and the variable $x_{\rm D}$ are defined by
\beal
\label{app:def_DnuD}
a
&=\frac{A_{21}}{4\pi \Delta\nu_{\rm D}}
\approx
\pot{8.61}{-4}\,\left[\frac{(1+z)}{1100\,\chi}\right]^{-1/2}
\\
\frac{\Delta\nu_{\rm D}}{\nu_{21}}
&=\sqrt{\frac{2\kB\Te}{m_{\rm H}c^2}}
\approx
\pot{2.35}{-5} \,\left[\frac{(1+z)}{1100\,\chi}\right]^{1/2}
\\
x_{\rm D}
&=\frac{\nu-\nu_{21}}{\Delta\nu_{\rm D}}.
\end{align}
\esub
Here $\nu_{21}\approx\pot{2.47}{15}\,$Hz is the Lyman $\alpha$ transition
frequency and $\chi=\Tg/\Te$.

The Voigt profile is normalized such that $\int_0^\infty
\frac{\varphi(\nu)}{4\pi}\id\nu\id\Omega\equiv 1$ and it has the well known
limiting cases
\beal
\label{app:Voigt_appr}
\phi(\nu)\approx
\begin{cases}
\frac{a}{\pi x_{\rm D}^2} &\text{for}\;|\xD|\gtrsim 10
\\
\mathcal{N}\,\frac{e^{-x_{\rm D}^2}}{\sqrt{\pi}}  &\text{for}\;\xD\sim 0
\end{cases}
\Abst{,} 
\end{align}
with $\mathcal{N}=e^{a^2} \Erfc(a)\approx 1-2a/\sqrt{\pi}+a^2$ for $a\ll1$ and
where $\Erfc(x)$ denotes the complementary error function.

In addition, on the red side of the resonance one can approximate the integral
$\chi=\int_0^{\nu}\phi(\nu') \id \nu'$ by
\beal
\label{app:Int_Wings_appr}
\chi_{\rm wings}
\approx
\pot{2.73}{-6}\,\left[\frac{(1+z)}{1100\,\chi}\right]^{-1/2} \left[\frac{\xD}{-100}\right]^{-1} 
\Abst{,} 
\end{align}
as long as $-\nu_{21}/\Delta\nu_{\rm D}\ll \xD \lesssim -10$. \changeF{This formula shows that only a very small fraction of photons is directly
emitted 
in the distant wings. 
However it is also known that escape of photons from the Doppler core is extremely
strongly suppressed. As a result the emission of the photons in the distant
wings should be considered carefully.
}

\subsubsection{Line emission term}
\label{sec:em_term}
With the definitions given \changeF{above,} the term for {\rm real} emission of photons
due the addition of {\it fresh electrons} to the 2p state can be written as
\beal
\label{app:coll_real_em}
\frac{1}{c}\left.\Abl{N_{\nu}}{t}\right|^{\rm em}_{\rm Ly-\alpha}
&=p_{\rm em}\,\frac{\phi(\nu)}{4\pi\,\Delta\nu_{\rm D}}\times
R^{+}_{\rm 2p}
\Abst{.} 
\end{align}
The emission probability, $p_{\rm em}$, is defined in Eq.~\eqref{app:p_em}.
Again we have neglected the factors related to stimulated emission.

For Eq.~\eqref{app:coll_real_em} we have assumed that the emission profile for
every new electron that was added to the 2p state is given by
Eq.~\eqref{app:Voigt}, regardless of whether the electron came from the continuum or from
some excited state.
In the absence of collisions (a very good approximation for the expanding
Universe) this is the standard approach, in which fresh electrons, i.e. those
that have {\it not} reached the 2p state by a line scattering event, lead to a
{\it natural excitation} of the 2p state \citep[e.g. see p.
  433][]{Mihalas1978}.
Note that this also implicitly means that any transition of electrons from the
2p state to higher levels \changeF{effectively} leads to a complete redistribution of photons in the
Lyman $\alpha$ line. 

%
One does expect some corrections related to these approximations, since even
for the real line emission process the history of the electron should matter
\citep[e.g. due to two-photon processes][]{Chluba2008a}. However, this problem
is beyond the scope of this paper.

\subsubsection{Line absorption term}
\label{sec:abs_term}
In the standard formulation \citep[e.g. see][p. 278]{Mihalas1978} the profile
for real line absorption is usually assumed to have the same shape as the
natural emission profile \eqref{app:Voigt}.
In this case, using the death probability $p_{\rm d}$, the term for real
line absorption reads
\beal
\label{app:coll_real_ab_incorr}
\frac{1}{c}\left.\Abl{N_{\nu}}{t}\right|^{\rm abs}_{\rm Ly-\alpha}
&=
p_{\rm d}\,h\nu_{\rm 21}\,B_{12}\, N_{1\rm s}\,\frac{\phi(\nu)}{4\pi\,\Delta\nu_{\rm D}}\,N_{\nu}
\Abst{.} 
\end{align}
Here $N_{\rm 1s}$ is the number density of hydrogen atoms in the ground state,
and $B_{12}$ is the Einstein coefficient for Lyman $\alpha$ absorption.
The idea behind Eq.~\eqref{app:coll_real_ab_incorr} is that only a fraction
$p_{\rm d}$ of the photons interacting with the Lyman $\alpha$ line really
undergo transitions to higher levels or the continuum, while most of the
interactions ($p_{\rm sc}\equiv p_{\rm em}=1-p_{\rm d}$) actually should be
considered as {\it resonance scattering} events.

More rigorously, using the principle of
detailed balance, instead of the standard absorption coefficient
$\alpha_{\nu}^{\rm st}=h\nu_{\rm 21}\,B_{12}\, N_{1\rm
  s}\,\frac{\phi(\nu)}{4\pi\,\Delta\nu_{\rm D}}$
\citep[e.g. see][p. 78]{Mihalas1978},
%
%
from Eq.~\eqref{app:coll_real_em}, also including the effect of stimulated
emission, one would deduce
$\alpha_{\nu}=\frac{\nu_{21}^2}{\nu^2}\,e^{h[\nu-\nu_{21}]/k\Tg}\times\alpha_{\nu}^{\rm
  st}$.
Although \changeF{especially the exponential term should} lead to significant
differences in the distant wings of the Lyman $\alpha$ line, we follow the
standard approximation and set
$\frac{\nu_{21}^2}{\nu^2}\,e^{h[\nu-\nu_{21}]/k\Tg}\sim 1$ in this expression.

It is clear that in the distant wings other corrections also will become very
important \citep[e.g. due to two-photon emission][]{Chluba2008a}, but a full
consideration of these aspects is beyond the scope of this paper.
However, in the standard formulation,
i.e. setting $\frac{\nu_{21}^2}{\nu^2}\,e^{h[\nu-\nu_{21}]/k\Tg}\sim 1$,
already at $|\xD|\gtrsim 100-1000$ a blackbody distribution is not exactly
conserved in full equilibrium. At the level of accuracy required in the
cosmological recombination problem this aspect will have to be resolved.

\subsubsection{Final line emission and absorption term}
\label{sec:em_abs_term}
With Eq.~\eqref{app:coll_real_em} and \eqref{app:coll_real_ab_incorr} one can
now write down the collision term for real line emission and absorption as
\bsub
\label{app:coll_real_emab}
\beal
\label{app:coll_real_emab_a}
\left.\mathcal{C}[N_{\nu}]\right|_{\rm e/a}
&=
p_{\rm em}\,\frac{\phi(\nu)}{4\pi\,\Delta\nu_{\rm D}}\,R^{+}_{\rm 2p}-p_{\rm d}\,\sigma_{\rm r}\, N_{1\rm s}\,\phi\,N_\nu
\nonumber\\[1mm]
&=
p_{\rm d}\,\sigma_{\rm r}\, N_{\rm 1s}\,\phi(\nu)
\left\{
N_{\rm em}-N_\nu\right\}
\\
\label{app:coll_real_emab_b}
N_{\rm em}&=\frac{p_{\rm em}}{\pdeath}\,\frac{R^{+}_{\rm 2p}}{h\nu_{21}\,B_{12}\,N_{\rm 1s}}
=\frac{2\nu_{21}^2}{c^2}\,\frac{g_{\rm 1s}}{g_{\rm 2p}}\,\frac{R^{+}_{2 \rm
    p}}{R^{-}_{2 \rm
    p}N_{\rm 1s}}.
\end{align}
\esub
Here we used the resonant scattering cross section
\beal
\label{app:sigma_r}
\sigma_{\rm r}
&
=\frac{h\nu_{21}}{4\pi}\,\frac{B_{12}}{\Delta\nu_{\rm D}}
\equiv
\frac{\pi e^2}{\me c}\,\frac{f_{12}}{\Delta\nu_{\rm D}}
\equiv
\frac{3\lambda^2_{21}a}{2}
\nonumber\\
&\approx\pot{1.91}{-13}{\rm cm^2}\left[\frac{(1+z)}{1100\,\chi}\right]^{-1/2},
\end{align}
and the Einstein relations $A_{21}=\frac{2 h\nu_{21}^3}{c^2}\,B_{21}$ and
$B_{21}=\frac{g_{\rm 1s}}{g_{\rm 2p}} B_{12}$, where $\lambda_{21}=c/\nu_{21}$
is the Lyman $\alpha$ wavelength, and $f_{12}$ is the absorption oscillator
strength of the Lyman $\alpha$ transition.
Note that $N_{\rm em}$ is only a function of redshift, but not frequency.
Furthermore, due to the factor $R^{+}_{\rm 2p}$ it depends on the solution for
the population of the higher levels.

\subsection{Transfer equation including line emission, line absorption and
  coherent scattering in the lab frame}
\label{sec:TEq_coh}
For coherent scattering in the lab frame no redistribution of photons over
frequency occurs. 
Using Eq.~\eqref{app:kin_eq_Nt_x_2} and \eqref{app:coll_real_emab}, the
time-dependent transfer equation therefore reads
\beal
\label{app:Eq_Nnu_coh}
\frac{1}{c}
\left.\pAb{\tilde{N}_x}{t}\right|_{x}
=p_{\rm d}\,\sigma_{\rm r}\, N_{\rm 1s}\,\phi(\nu)
\left\{
\tilde{N}_{\rm em}-\tilde{N}_x\right\}
\Abst{,} 
\end{align}
with $\nu=x\,(1+z)$, and $\tilde{N}_{\rm em}=N_{\rm em}/(1+z)^2$, where $N_{\rm
  em}$ is defined by Eq.~\eqref{app:coll_real_emab_b}.
For the initial condition $\tilde{N}_{x}(z_{\rm s})=\tilde{N}^{\rm pl}_x$,
where $z_{\rm s}$ is a redshift well before the epoch of hydrogen cosmological
recombination, this equation formally has the simple solution
\beal
\label{app:kin_abs_em_Sol}
\tilde{N}_x(z)&=\tilde{N}^{\rm pl}_x-
\int^{z_{\rm s}}_z [\tilde{N}_{\rm em}(z')-\tilde{N}^{\rm pl}_x]
\,
\partial_{z'}e^{-\tau_{\rm abs}(x, z', z)}\id z'.
\end{align}
Here $\tau_{\rm abs}(x, z', z)$ is defined by
\bsub
\label{app:tau_abs_all}
\beal
\label{app:tau_abs}
\tau_{\rm abs}(x, z', z)
&=\int_{z}^{z'}
p_{\rm d}\,\frac{c\,\sigma_{\rm r}\,N_{\rm 1s}}{H (1+\tilde{z})}\,\phi(x[1+\tilde{z}])\id \tilde{z}
\\
\label{app:tau_abs_nut}
&=\int_{\nu}^{\nu'}
p_{\rm d}\,\frac{c\,\sigma_{\rm r}\,N_{\rm 1s}}{H}\,\phi(\tilde{\nu})\,\frac{\id \tilde{\nu}}{\tilde{\nu}}.
\end{align}
\esub
In Eq.~\eqref{app:tau_abs_nut} we have used the substitution
$\tilde{\nu}=x(1+\tilde{z})$, so that the current redshift can be found from
$1+\tilde{z}=\tilde{\nu}(1+z)/\nu$.
Note that in the given set of variables $\tilde{N}^{\rm pl}_x$ does not \changeF{explicitly} depend
on redshift, so we omitted it in the notation.

Returning to physical coordinates one can finally write
\bsub
\label{app:kin_abs_em_Sol_phys}
\beal
\label{app:kin_abs_em_Sol_phys_a}
\Delta N^{\rm coh}_{\nu}(z)&\!=\![N_{\rm em}(z)-N^{\rm pl}_{\nu_{21}}(z)]\!\!\int_{z_{\rm s}}^z \!\!\Theta^{\rm coh}(z')
\,
\partial_{z'}e^{-\tau_{\rm abs}(\nu, z', z)}\id z',
\\
\label{app:kin_abs_em_Sol_phys_b}
\Theta^{\rm coh}(z')
&\!=\!\frac{\tilde{N}_{\rm em}(z')-\tilde{N}^{\rm pl}_x}{\tilde{N}_{\rm em}(z)-\tilde{N}^{\rm pl}_{x_{21}}}
\end{align}
\esub
\changeF{where $x_{21}=\nu_{21}/(1+z)$ and} in expression \eqref{app:tau_abs_all} for $\tau_{\rm abs}$ one should use
$x=\nu/(1+z)$.

\subsection{Transfer equation including line emission, line absorption and
  complete redistribution}
\label{sec:TEq_CR}
In the case of complete redistribution one has to add the term \citep[see
e.g.][]{Mihalas1978}
\beal
\label{app:scatt_CR}
\left.\mathcal{C}[N_{\nu}]\right|_{\rm r}
&\approx
p_{\rm sc}\,\sigma_{\rm r} N_{1\rm s} \phi(\nu)
\left[\bar{N}-N_{\nu}\right]
\end{align}
to Eq.~\eqref{app:Eq_Nnu_coh}.
Here $\bar{N}=\int\frac{\varphi(\nu)}{4\pi} N_{\nu} \id\nu\id\Omega$.
This then yields
\beal
\label{app:Eq_Nnu_CR}
\frac{1}{c}
\left.\pAb{\tilde{N}_x}{t}\right|_x
=\sigma_{\rm r}\, N_{\rm 1s}\,\phi(\nu)
\left\{
\tilde{N}^{\rm cr}_{\rm em}-\tilde{N}_x\right\}
\Abst{.} 
\end{align}
with $\tilde{N}^{\rm cr}_{\rm em}=\pdeath\tilde{N}_{\rm
  em}+\psc\tilde{\bar{N}}$, and $\tilde{\bar{N}}=\bar{N}/(1+z)^2$.
Although complete redistribution is not appropriate for conditions valid in
the expanding Universe (practically no collisions), it is used many times in
the literature, in particular for the derivation of the Sobolev escape
probability.
One should mention that in this approach it is assumed that even those photons
scattering in the very distant red wing of the Lyman $\alpha$ resonance can
directly return to the line center in one scattering event. With Doppler
redistribution, which is described by so-call type-II redistribution
\citep{Hummer1962}, this is only possible after many scatterings (if at all),
or by a real line absorption event.

Comparing Eq.~\eqref{app:Eq_Nnu_CR} with Eq.~\eqref{app:Eq_Nnu_coh}, in
physical coordinates one can directly write down the solution as
\bsub
\label{app:kin_abs_em_Sol_CR_phys}
\beal
\label{app:kin_abs_em_Sol_CR_phys_a}
\Delta N^{\rm cr}_{\nu}(z)&=[N^{\rm cr}_{\rm em}(z)-N^{\rm pl}_{\nu_{21}}(z)]\,\int_{z_{\rm s}}^z \Theta^{\rm cr}(z')
\,
\partial_{z'}e^{-\tau_{\rm cr}(\nu, z', z)}\id z',
\\
\label{app:kin_abs_em_Sol_CR_phys_b}
\Theta^{\rm cr}(z')
&=\frac{\tilde{N}^{\rm cr}_{\rm em}(z')-\tilde{N}^{\rm pl}_x}{\tilde{N}^{\rm cr}_{\rm em}(z)-\tilde{N}^{\rm pl}_{x_{21}}}
\end{align}
\esub
where $\tau_{\rm cr}\equiv \tau_{\rm abs}|_{\pdeath=1}$. It is clear that
$\tau_{\rm cr}\gg \tau_{\rm abs}$ since $\pdeath\ll 1$ during cosmological
recombination (see Fig.~\ref{fig:p_d}).

\section{The Lyman $\alpha$ escape problem and results for the escape probabilities}
\label{sec:Ly-a-problem}
In order to solve the cosmological recombination problem, the usual way is to
separate the evolution of the photon field from the evolution of the matter,
in particular the populations of the different energy states inside the
hydrogen atom.
This is normally achieved using the Sobolev approximation for the optically
thick Lyman series in order to define the mean intensity of photons supporting
the $n$p-state at a given time, and leads to the definition of the {\it
  Sobolev escape probability}. 
In this Section we explain the details of this approximation and
compare it with other cases that can be solved analytically.

\subsection{The Lyman $\alpha$ net rate}
\label{sec:Ly-trans}
\label{sec:net_rate}
The net change of the number density of electrons in the 2p level via the
Lyman $\alpha$ channel is given by
\beal
\label{eq:dN2pdt_g}
\left.\Abl{N_{\rm 2p}}{t}\right|_{\rm Ly-\alpha}
&=A_{\rm 21}\frac{g_{\rm  2p}}{g_{\rm 1s}}\,N_{\rm 1s} 
\,\bar{n}-A_{\rm 21}(1+\bar{n})\,N_{\rm 2p}
\nonumber\\
&=A_{\rm 21}(1+\bar{n})\,N_{\rm 2p}
\left[\frac{g_{\rm 2p}}{g_{\rm 1s}}\,\frac{N_{\rm 1s}}{N_{\rm 2p}}\,\frac{\bar{n}}{1+\bar{n}}-1 \right]
\end{align}
where $A_{\rm 21}=\pot{6.27}{8}\,\text{s}^{-1}$ is the spontaneous 2p-1s
transition rate, $N_i$ denotes the number density of electrons in level
$i$. Furthermore we made use of the Einstein relations, and defined
$\bar{n}=\frac{c^2 \bar{J}}{2h\nu_{21}^3}$ with
\beal
\label{eq:nbar_Ly-a}
\bar{J}=\int_0^\infty \frac{\varphi(\nu)}{4\pi}\,I_{\nu} \id\nu\id\Omega.
\end{align}
According to the textbook
derivations\footnote{Also in the derivation of the Einstein relations this
  approximation is normally applied.}
$\bar{J}\approx J(\nu_{21})\approx h\nu_{21}\bar{N}\approx
h\nu_{21}\int\frac{\varphi(\nu)}{4\pi} N_{\nu} \id\nu\id\Omega\approx
\frac{2h\nu_{21}^3}{c^2}\bar{n}$ with very high accuracy, since $\varphi(\nu)$
is so sharply peaked at $\nu\sim \nu_{21}$.

Defining the {\it line occupation number}
\beal
\label{eq:def_nL}
n_{\rm L}&=\frac{1}{\frac{g_{\rm 2p}}{g_{\rm 1s}}\,\frac{N_{\rm 1s}}{N_{\rm 2p}}-1}
\approx\frac{g_{\rm 1s}}{g_{\rm 2p}}\,\frac{N_{\rm 2p}}{N_{\rm 1s}}
\end{align}
Eq.~\eqref{eq:dN2pdt_g} can be cast into the form
\beal
\label{app:Ly-trans_nL}
\left.\Abl{N_{\rm 2p}}{t}\right|_{\rm Ly-\alpha}
&\approx
A_{\rm 21}\,\frac{g_{\rm 2p}}{g_{\rm 1s}}\,N_{\rm 1s}\,\Delta\bar{n}^{\rm L},
\end{align}
where we have introduced $\Delta\bar{n}^{\rm L}=\bar{n}-\nL$, for which we
will now discuss different approximations below.

\subsection{Escape probability within the Sobolev approximation}
\label{sec:Sobolev}
The aim is now to determine the solution for the mean occupation number in the
Lyman $\alpha$ resonance using the Sobolev approximation.
The two key assumptions for its derivation are (i) {\it quasi-stationary}
evolution of the photon field and (ii) that every resonance scattering leads
to a {\it complete redistribution} of photons over the whole Lyman $\alpha$
line profile.
With these assumptions we can obtain the solution for the spectral distortion
at redshift $z$ using the results of Sect.~\ref{sec:TEq_CR}.
Under quasi-stationary conditions one can simply set\footnote{\changeF{In fact this approximation not only implies quasi-stationarity, i.e. $\tilde{N}^{\rm cr}_{\rm em}(z')=\tilde{N}^{\rm cr}_{\rm em}(z)$, but also that one can use $\tilde{N}^{\rm pl}_x\approx \tilde{N}^{\rm pl}_{x_{21}}$.}} $\Theta^{\rm cr}(z')=1$
in Eq.~\eqref{app:kin_abs_em_Sol_CR_phys}, and for the absorption optical
depth, $\tau_{\rm cr}$, one has
\beal
\label{app:tau_abs_nut_appr}
\tau^{\rm qs}_{\rm cr}(\nu, z', z)
&\approx \tau_{\rm S}(z)\int_{\nu}^{\nu'} \varphi(\tilde{\nu})\id \tilde{\nu},
\end{align}
where $\varphi(\nu)=\varphi(\nu, z)$ is given by Eq. \eqref{app:Voigt}, and
$\nu'=\nu\,\frac{1+z'}{1+z}$.
Furthermore we introduced the {\it Sobolev} optical depth of the
Lyman $\alpha$ line
\beal
\label{app:tau_S}
\tau_{\rm S}=\frac{c\,\sigma_{\rm r}\,N_{\rm 1s}}{H}\,\frac{\Delta\nu_{\rm D}}{\nu_{21}}
=\frac{g_{\rm 2p}}{g_{\rm 1s}}\,\frac{A_{21} \lambda_{21}^3}
{8\pi H}\,N_{\rm 1s}
\end{align}
with wavelength $\lambda_{21}=c/\nu_{21}$. 
For Eq. \eqref{app:tau_abs_nut_appr} we have assumed that $\tau_{\rm S}$ does
not change significantly between $z'$ and $z$. Also we have neglected the
variation of $1/\tilde{\nu}$ in comparison with $\varphi(\nu)$, and set
$1/\tilde{\nu}\approx 1/\nu_{21}$.
This approximation is normally applied in the literature and computation of
the recombination history.

From Eq.~\eqref{app:kin_abs_em_Sol_CR_phys} \changeF{with $\Theta^{\rm cr}=1$} one then obtains
\beal
\label{app:kin_abs_em_Sol_CR_phys_QS}
\Delta N^{\rm cr, qs}_{\nu}(z)&=[N^{\rm cr}_{\rm em}-N^{\rm pl}_{\nu_{21}}]\,
\left[1-e^{-\tau_{\rm S}}e^{\tau_{\rm S}\chi}\right],
\end{align}
with $\chi(\nu)=\int_0^\nu \varphi(\nu') \id\nu'$.

\begin{figure}
\centering 
\includegraphics[width=0.9\columnwidth]
{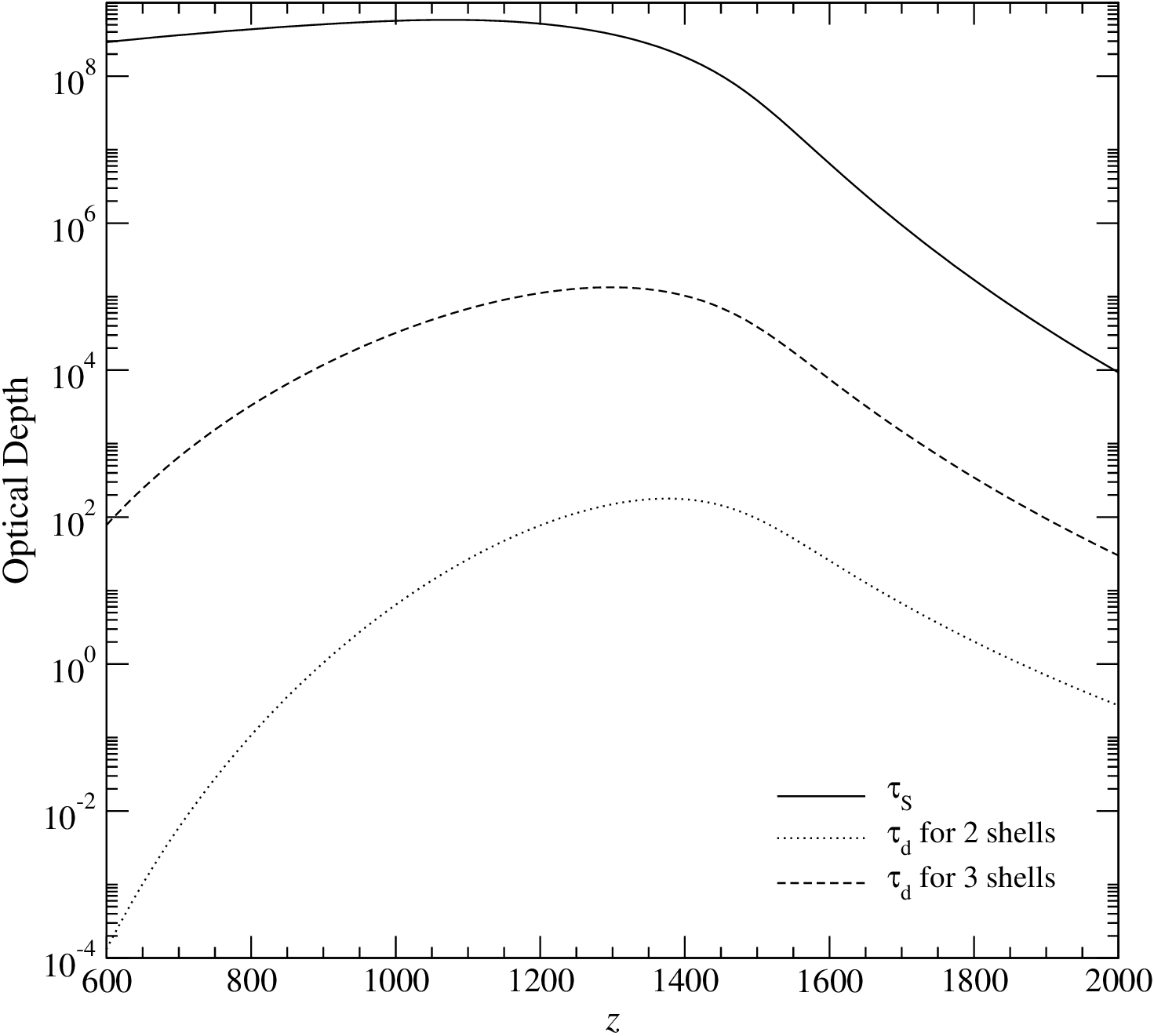}
\caption
{Comparison of the Sobolev optical depth, $\tau_{\rm S}$, as defined by
  Eq.~\eqref{app:tau_S}, and the total absorption optical depth, $\tau_{\rm
  d}=\pdeath\tauS$, for the 2- and 3 shell case.
To compute $\tauS$ we used the {\sc Recfast} solution for $N_{\rm 1s}$. }
\label{fig:tau_d_S}
\end{figure}

\begin{figure}
\centering 
\includegraphics[width=0.90\columnwidth]
{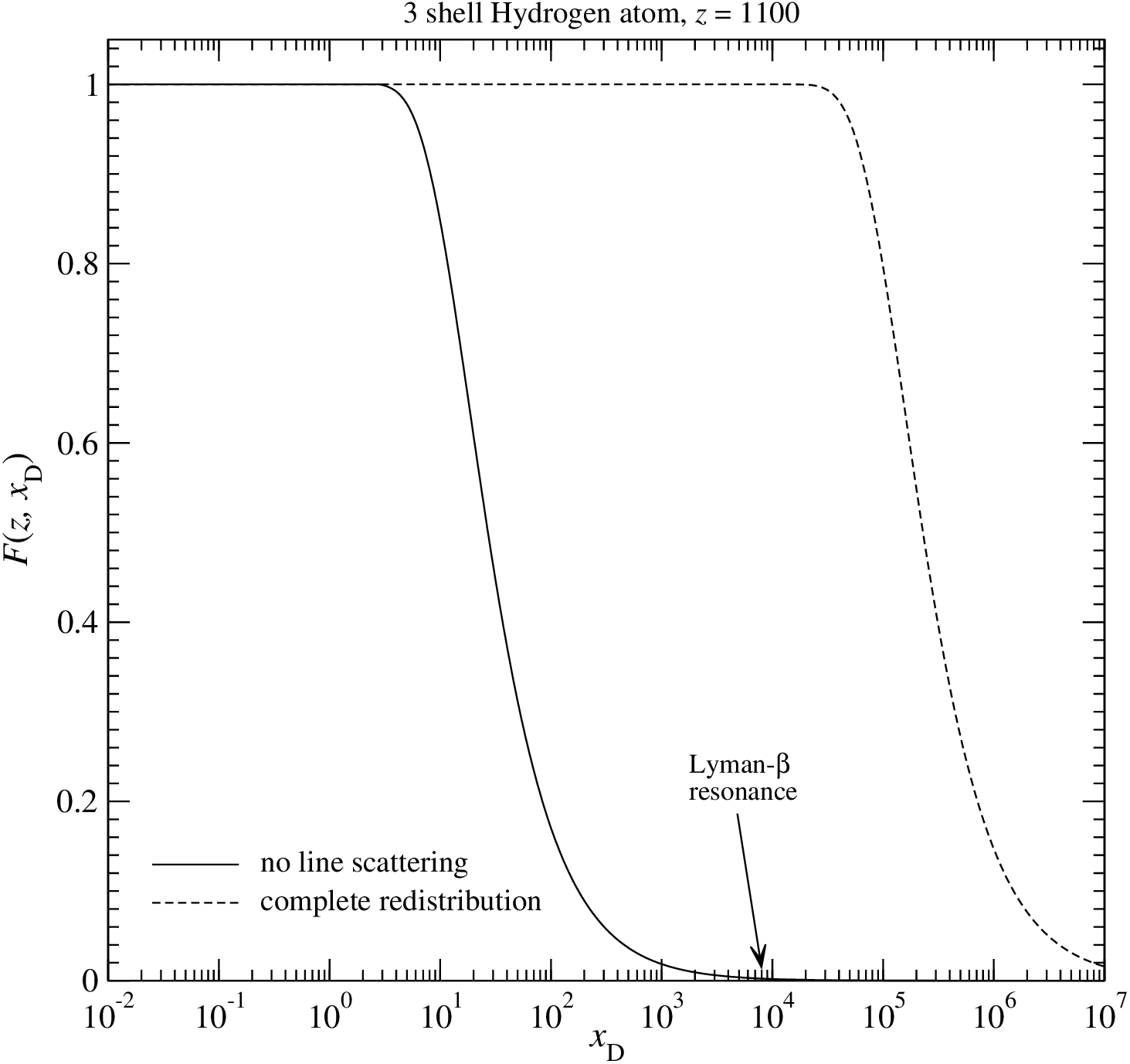}
\caption
{Spectral behavior of the solutions in the quasi-stationary approximation at
  redshift $z=1100$. We normalized the distortion to unity at the
  Lyman $\alpha$ frequency, i.e. $F\equiv\Delta N_\nu(z)/\Delta
  N_{\nu_{21}}(z)$. The solid line shows the result obtained in the no
  line scattering approximation, while the dashed line represents the solution
  for the complete redistribution case. We also indicated the position of the
  Lyman $\beta$ resonance at that time.}
\label{fig:spec_QS}
\end{figure}
\subsubsection{Spectral characteristics of the solution}
As can be seen in Fig.~\ref{fig:tau_d_S}, during hydrogen recombination
$\tauS\gg 1$. 
\changeF{According to
Eq.~\eqref{app:kin_abs_em_Sol_CR_phys_QS} the photon distribution} therefore
varies strongly close to\footnote{There the value of $\Delta N^{\rm cr,
    qs}_{\nu}(z)$ has decreased by a factor of 2 compared to the line
  center.} $\tau_{\rm S}[1-\chi]\sim \ln 2$, while it is basically identical to
unity\footnote{With relative accuracy better than $\sim e^{-\tauS/2}$.} at $\xD\leq
0$.
Using the wing-expansion \eqref{eq:Int_Wings_appr} of the Voigt profile one
therefore finds that this happens at a distance of about
\beal
\label{eq:xD_Sobolev}
x_{\rm D}\approx \frac{a}{\pi}\,\frac{\tauS}{\ln 2}.
\end{align}
At $z\sim 1100$ one has $a\sim\pot{8.6}{-4}$ and $\tauS\sim \pot{5.8}{8}$,
such that $x_{\rm D}\sim \pot{2.3}{5}$.
This shows that in the complete redistribution approximation the photon
distribution is in full equilibrium with the value at the line center up to
extremely large distances on the blue side of the line center (see
Fig.~\ref{fig:spec_QS}).

Physically this type of redistribution does not describe the problem very
accurately, and a much more realistic solution is obtained using the
case of coherent scattering in the lab frame (see Sect.~\ref{sec:ncoh}).
For example, if we consider the position of the Lyman $\beta$ line at redshift
$z=1100$, then in Doppler units of the Lyman $\alpha$ line one finds $x^{\rm
  Ly_\beta}_{\rm D}\sim 8000$.
The variation of the photon distribution, which is important for the value of
the escape probability (see below), occurs far beyond this value. In
fact, $x_{\rm D}\sim \pot{2.3}{5}$ corresponds to about 2 times the
Lyman $\alpha$ frequency, or $1.5$ times the ionization energy of the hydrogen
atom.
The Sobolev optical depth, $\tauS$, for conditions during recombination is
simply so large that the approximation of complete redistribution \changeF{becomes
unphysical}.

Furthermore, at such large distances it is even questionable as to why one should be
able to neglect variations of the blackbody distribution, or the factor of
$1/\tilde{\nu}$ in the definition of $\tau_{\rm abs}$.
However, such an approximation is necessary to obtain the expression for the
Sobolev escape probability.
Obviously other corrections (e.g. related to two-photon processes, or the
imbalance in the emission and absorption coefficient as mentioned in
Sect.~\ref{sec:abs_term}) will become important and {\it even necessary} to
correct for these physical discrepancies. However, as we will see below,
in spite of all these problems the Sobolev approximation at the level of $\sim
10\%$ provides the correct answer for the escape probability, a fact that is
very surprising.

\subsubsection{Mean occupation number in the Lyman $\alpha$ and the Sobolev escape probability}
To obtain the mean photon occupation number in the Lyman $\alpha$ line we
multiply \eqref{app:kin_abs_em_Sol_CR_phys_QS} by $\varphi$ and integrate over
$\nu$. This then yields
\bsub
\label{eq:Delta_n_CR}
\beal
\label{eq:Delta_n_CR_a}
\Delta\bar{n} 
&=\bar{n}-\bar{n}^{\rm pl}
\approx
\left\{ p_{\rm em}\,\bar{n}+p_{\rm d}\,n_{\rm em}-n_{\nu_{21}}^{\rm pl}\right\} \left[1-P_{\rm S}\right]
\\
\label{eq:Delta_n_CR_b}
P_{\rm S}
&=
1-\int_0^1 \id\chi \,e^{-\tauS\,[1-\chi]}
=
\frac{1-e^{-\tauS}}{\tauS}\stackrel{\stackrel{\tau_{\rm S}\gg 1}{\downarrow}}{\approx}\frac{1}{\tau_{\rm S}},
\end{align}
\esub
where we again have neglected the variation of $n^{\rm pl}$ over the line
profile.
Here $P_{\rm S}$ denotes the standard Sobolev escape probability (see
Fig.~\ref{fig:p_d} for its redshift dependence).
After some rearrangement and with $n_{\nu_{21}}^{\rm
  pl}\approx \bar{n}^{\rm pl}$, Eq.~\eqref{eq:Delta_n_CR_a} can be cast into the final form
\bsub
\label{eq:nb_CR}
\beal
\label{eq:nb_CR_a}
\bar{n}_{\rm cr}&=
\bar{n}^{\rm pl}+\frac{p_{\rm d}(1-P_{\rm S})}{p_{\rm d}+p_{\rm sc}P_{\rm S}}\,[n_{\rm em}-\bar{n}^{\rm pl}]
\\
\label{eq:nb_CR_b}
&\equiv n_{\rm em}-\frac{P_{\rm S}}{p_{\rm d}+p_{\rm sc}P_{\rm S}}\,[n_{\rm em}-\bar{n}^{\rm pl}]
\\
\label{eq:nb_CR_c}
&\!\!\!\!\!\stackrel{\stackrel{\pdeath\gg p_{\rm sc}P_{\rm S}}{\downarrow}}{\approx}
n_{\rm em}-\frac{P_{\rm S}}{p_{\rm d}}\,[n_{\rm em}-\bar{n}^{\rm pl}].
\end{align}
\esub
A solution similar to Eq.~\eqref{eq:nb_CR} was also given and discussed in
\citet{Hummer1992}.

\subsubsection{Relation to the expression which is normally used in multi-level
  recombination codes}
But how does Eq.~\eqref{eq:nb_CR} actually relate to the expression
\beal
\label{eq:Nb_Seager}
\bar{n}_{\rm st}&\approx n_{\rm L}-P_{\rm S}\,[n_{\rm L}-\bar{n}^{\rm pl}]
\equiv\bar{n}^{\rm pl}+[n_{\rm L}-\bar{n}^{\rm pl}](1-P_{\rm S})
\end{align}
that is normally used \citep[cf.][]{Seager2000} in computations of the
hydrogen recombination problem?
To understand this connection the key ingredient is the quasi-stationary
solution for the 2p population.
In fact this approximation should {\it always} be possible, even if the
spectral evolution is non-stationary, simply because the re-adjustment of the
2p population after some changes in the spectrum is so fast.

With Eqs.~\eqref{eq:dN2pdt} and \eqref{app:Ly-trans_nL}, the rate equation
governing the time evolution of the 2p state can be cast in the form
\beal
\label{eq:dX_2p_dt_CR}
N_{\rm H} \Abl{X_{\rm 2p}}{t}
&=R^{+}_{2\rm p}
-N_{\rm 2p}\,R^{-}_{2\rm p}
+A_{21}\,N_{\rm 2p}\,\left[\frac{\bar{n}}{n_{\rm L}}-1\right],
\end{align}
where we directly neglected induced terms, \changeF{and introduced $X_{\rm 2p}=N_{\rm 2p}/N_{\rm H}$, where $N_{\rm H}$ denotes the total number density of hydrogen nuclei}. 
With $\Abl{X_{\rm 2p}}{t}\approx 0$ one readily finds
\beal
\label{eq:N2p_QS}
N_{\rm 2p}^{\rm qs}\approx \frac{R^{+}_{2\rm p}}{A_{21}+R^{-}_{2\rm p}-A_{21}\frac{\bar{n}}{n_{\rm L}}}
=\frac{\pdeath \,n_{\rm L}}{n_{\rm L}-p_{\rm em}\bar{n}}\,\frac{R^{+}_{2\rm p}}{R^{-}_{2\rm p}}.
\end{align}
Inserting this into Eq. \eqref{app:coll_real_emab_b} therefore yields
\beal
\label{eq:N_em_QS}
n_{\rm L}\approx p_{\rm em}\,\bar{n}+p_{\rm d}\,n_{\rm em}.
\end{align}
If we now use this in Eq. \eqref{eq:Delta_n_CR} one immediately finds
$\bar{n}_{\rm cr}\equiv \bar{n}_{\rm st}$.
Therefore the result \eqref{eq:nb_CR} is completely equivalent to \changeF{Eq.~\eqref{eq:Nb_Seager}}.

However, in one case factors are expressed in terms of $n_{\rm em}$, while in
the other case $n_{\rm L}$ is used.
From Eq.~\eqref{eq:nb_CR_b} and \eqref{eq:Nb_Seager} with $\bar{n}_{\rm
  cr}\equiv \bar{n}_{\rm st}$ one can easily show
\beal
\label{eq:N_em_NL_relation}
n_{\rm em}-\bar{n}^{\rm pl}=\left(1+\frac{\psc P_{\rm S}}{\pdeath}\right)[n_{\rm L}-\bar{n}^{\rm pl}].
\end{align}
This implies that $n_{\rm em}\approx n_{\rm L}$. However, since in the
Lyman $\alpha$ rate, \changeF{Eq.~\eqref{app:Ly-trans_nL}}, the main term ($\sim
n_{\rm L}$) cancels, the small difference $n_{\rm em}-n_{\rm L} = \frac{\psc
  P_{\rm S}}{\pdeath}[n_{\rm L}-\bar{n}^{\rm pl}]$ cannot be neglected.
Replacing the first $n_{\rm em}$ in Eq.~\eqref{eq:nb_CR_c} with this
expression and neglecting terms $\mathcal{O}(P_{\rm S}/\pdeath)^2$ one can
directly recover Eq.~\eqref{eq:Nb_Seager}. However, here it is important that
in Eq.~\eqref{eq:nb_CR_c} there is partial cancellation of terms
$\mathcal{O}(P_{\rm S}/\pdeath)$ from the first and second $n_{\rm em}$,
leaving a much smaller residual $\sim P_{\rm S}[n_{\rm L}-\bar{n}^{\rm pl}]$.

\subsection{Escape probability for the case of coherent scattering}
\label{sec:ncoh}
In the absence of line scattering, or equivalently for {\it coherent
  scattering} in the lab frame, the solution of the transfer equation is given
by Eq.~\eqref{app:kin_abs_em_Sol_phys}. Under quasi-stationary conditions \changeF{(and with $\tilde{N}^{\rm pl}_x\approx \tilde{N}^{\rm pl}_{x_{21}}$)} one
again has $\Theta^{\rm coh}(z')=1$, and also it is possible to use $\tau_{\rm
  abs}^{\rm qs}\approx \pdeath\,\tau_{\rm cr}^{\rm qs}$, where $\tau_{\rm
  cr}^{\rm qs}$ is defined by Eq.~\eqref{app:tau_abs_nut_appr}.
Then one can write
\beal
\label{app:kin_abs_em_Sol_coh_phys_QS}
\Delta N^{\rm coh, qs}_{\nu}(z)&=[N_{\rm em}-N^{\rm pl}_{\nu_{21}}]\,\left[1-e^{-\tau_{\rm d}}e^{\tau_{\rm d}\chi}\right],
\end{align}
with $\tau_{\rm d}=\pdeath\tauS$.

\subsubsection{Spectral characteristics of the solution}
Looking at Fig.~\ref{fig:tau_d_S} it is clear that $\tau_{\rm d}\ll\tauS$ at
all relevant redshifts, so that $\Delta N^{\rm coh, qs}_{\nu}(z)$ should
change strongly much closer to the line center than in the complete
redistribution case, Eq. \eqref{app:kin_abs_em_Sol_CR_phys_QS}.
If we again want to estimate where the photon distribution
\eqref{app:kin_abs_em_Sol_coh_phys_QS} varies most rapidly, assuming that this
happens in the blue wing of the Lyman $\alpha$ line, we can find
\beal
\label{eq:xD_Sobolev_pd}
x_{\rm D}\approx \frac{a}{\pi}\,\frac{\pdeath\tau_{\rm S}}{\ln 2}.
\end{align}
At $z\sim 1100$ one has $\pdeath\sim 10^{-4}$, such that with
Eq.~\eqref{eq:xD_Sobolev} one has $x_{\rm D}\sim 30$ (see Fig.~\ref{fig:spec_QS}).
This result is much closer to the solution that would be obtained when using
the more realistic type-II redistribution for the Lyman $\alpha$ resonance
scattering process, where in the quasi-stationary approximation the photon
distribution for typical conditions in our Universe strongly varies at
distances of a few hundred Doppler width \citep[e.g. see][]{RybickiDell94}.
We already checked this point and found very similar results
\citep{Chlubaprep}.

\subsubsection{Mean occupation number in the Lyman $\alpha$ and the escape probability}
With \eqref{app:kin_abs_em_Sol_coh_phys_QS} and the same simplifications that
were mentioned above in connection with Eq.~\eqref{eq:nb_CR} one then obtains
\bsub
\label{app:nb_coh}
\beal
\label{app:nb_coh_a}
\bar{n}_{\rm coh}&=n_{\rm em}-P_{\rm d}[n_{\rm em}-\bar{n}^{\rm pl}]
\\
\label{app:Pd}
P_{\rm d}&=\frac{1-e^{-\tau_{\rm d}}}{\tau_{\rm d}}
\stackrel{\stackrel{\tau_{\rm d}\gg 1}{\downarrow}}{\approx}\frac{1}{\pdeath\tau_{\rm S}}\approx\frac{P_{\rm S}}{\pdeath}.
\end{align}
\esub
Note that $P_{\rm d}$ is very similar to the standard Sobolev escape
probability, $P_{\rm S}=[1-e^{-\tau_{\rm S}}]/\tau_{\rm S}$, with the only
difference that in general $\tau_{\rm d}\leq \tau_{\rm S}$ and hence $P_{\rm
  d}\geq P_{\rm S}$ (cf. Fig.~\ref{fig:p_d} and \ref{fig:tau_d_S}).  Also one
can directly see that in lowest order Eq.~\eqref{app:nb_coh} is identical to
Eq.~\eqref{eq:nb_CR_c}.
This already suggests that in both the complete redistribution and the no
line scattering approximation the answer for $\bar{n}$ is nearly the same,
with differences of the order $\mathcal{O}(P_{\rm S}/\pdeath)^2$.
%

\begin{figure}
\centering 
\includegraphics[width=0.90\columnwidth]
{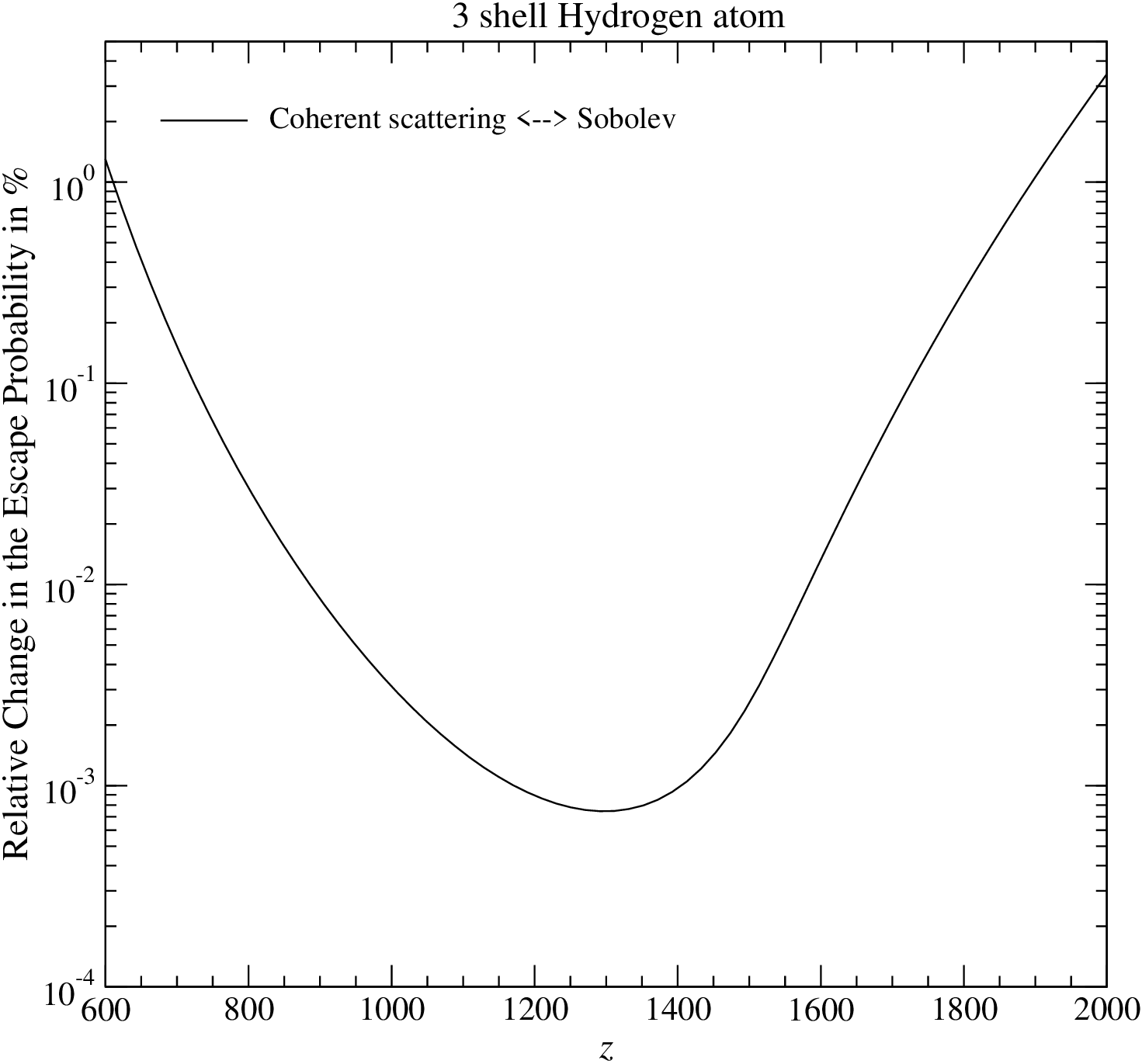}
\caption
{Relative difference between the escape probability $P^{\rm L}_{\rm
    d}=\frac{p_{\rm d}\,P_{\rm d}}{1-p_{\rm em}\,P_{\rm d}}$ and the standard
  Sobolev escape probability, $P_{\rm S}$.
The death probability for a 3 shell hydrogen atom was used.  
}
\label{fig:DP.P_d.P_S}
\end{figure}
Again using the quasi-stationary solution for the 2p population, we can
replace $n_{\rm em}$ applying the expression Eq.~\eqref{eq:N_em_QS}. Then
solving for $\bar{n}$ one finds
\beal
\label{app:nb_QS_coh}
\bar{n}_{\rm coh}&\approx n_{\rm L}-\frac{p_{\rm d}\,P_{\rm d}}{1-p_{\rm em}\,P_{\rm d}}[n_{\rm L}-\bar{n}^{\rm Pl}].
\end{align}
Comparing this with the standard form Eq.~\eqref{eq:Nb_Seager}, it is again
clear that for $p_{\rm em}\,P_{\rm d}\ll 1$ and $\tau_{\rm S}\gg \tau_{\rm
  d}\gg 1$ one has $\bar{n}_{\rm coh}\approx \bar{n}_{\rm st}$.
Looking at Fig.~\ref{fig:p_d} shows that this should be the case at most of
the redshifts relevant for the recombination of hydrogen, and that the
differences between the complete redistribution and no scattering case should
not exceed the level of $\sim 10^{-3}$ in the redshift range $800\lesssim z
\lesssim 1600$. 

In Fig.~\ref{fig:DP.P_d.P_S} we present a more detailed comparison and indeed
find practically no important difference to the standard Sobolev case.
This result is somehow surprising, since the assumption of complete
redistribution leads to a totally different (and physically unrealistic)
solution for the photon distribution. Still, the final result is comparable.
This is due to the fact that the changes in the shape of the photon
distribution are compensated by changes in the amplitude of the spectrum close
to the line center, as already explained in connection with
Eq.~\eqref{eq:N_em_NL_relation}.
%

%
Note that in the case of 2 shells the differences would be much greater, since
one can find $P_{\rm d}\sim 1$ at redshifts relevant for recombination (see
Fig.~\ref{fig:p_d}).
This again shows that one has to include at least 3 shells in the computation,
in order to obtain meaningful results.

\subsubsection{Escape probability in the limit $\pdeath\rightarrow 1$}
It is also illustrative to look at the solution in the limit
$\pdeath\rightarrow 1$. Physically, in the current formulation of the problem
this should give the same answer as in the approximation of complete
redistribution. This is because for $\pdeath=1$ every electron entering the
2p level via the Lyman $\alpha$ channel will pass through the continuum or
some higher shell, where it will forget its history. It will be replaced by
another fresh electron, with a natural line profile, as in the complete
redistribution approximation for a line scattering event.

From Eq.~\eqref{app:nb_QS_coh}, with $\pdeath=1$, $\psc=0$ and $P_{\rm
  d}\equiv P_{\rm S}$, it is quite obvious that $\bar{n}_{\rm coh}\equiv
\bar{n}_{\rm cr}$, but can one also see this directly from
Eq.~\eqref{app:nb_coh_a}, which in the first place only leads to $\bar{n}_{\rm
  coh}=n_{\rm em}-P_{\rm S}[n_{\rm em}-\bar{n}^{\rm pl}]$.
Here apparently $n_{\rm em}\equiv n_{\rm L}$, a result that indeed can be
confirmed with Eq.~\eqref{eq:N_em_QS}, so that also $\bar{n}_{\rm coh}\equiv
\bar{n}_{\rm cr}$ follows.

\subsubsection{Until what distance from the line center is the shape of the photon
  distribution important?}
\label{sec:distance}
The escape probability, $P_{\rm d}$, was obtained from the integral over the
Lyman $\alpha$ line profile. If \changeF{we only integrate} up to some
frequency $\nu_{\rm m}$, then one has
\beal
\label{eq:P_d_max}
P_{\rm d}(\nu_{\rm m})
&=\int_0^{\nu_{\rm m}}
\varphi(\nu)\,e^{-\tau_{\rm d}}e^{\tau_{\rm d}\chi(\nu)}\id\nu
=\frac{e^{-\tau_{\rm d}[1-\chi_{\rm m}]}-e^{-\tau_{\rm d}}}{\tau_{\rm d}}
\nonumber\\
&\approx\frac{e^{-\tau_{\rm d}[1-\chi_{\rm m}]}}{\tau_{\rm d}},
\end{align}
with $\chi_{\rm m}=\chi(\nu_{\rm m})$. Therefore the relative difference \changeF{in the}
value of $P_{\rm d}$ is given by
\beal
\label{eq:DP_d_max}
\frac{\Delta P_{\rm d}(\nu_{\rm m})}{P_{\rm d}}
&=\frac{P_{\rm d}(\nu_{\rm m})-P_{\rm d}}{P_{\rm d}}
\nonumber\\
&\approx e^{-\tau_{\rm d}[1-\chi_{\rm m}]}-1
\stackrel{\stackrel{\tau_{\rm d}[1-\chi_{\rm m}]\ll 1}{\downarrow}}{\approx}-\tau_{\rm d}[1-\chi_{\rm m}].
\end{align}
If we assume that $\nu_{\rm m}>\nu_{21}$ is already in the damping wing then
with the approximation Eq.~\eqref{eq:Int_Wings_appr} of the Voigt profile one
obtains
\beal
\label{eq:DP_d_max_appr}
\frac{\Delta P_{\rm d}(\nu_{\rm m})}{P_{\rm d}}
&\approx -\frac{a}{\pi} \frac{\tau_{\rm d}}{\xD}.
\end{align}
At $z=1100$ this yields $\frac{\Delta P_{\rm d}(\nu_{\rm m})}{P_{\rm d}}=-16\%
\left[\frac{\xD}{100}\right]^{-1}$, so that in the no-scattering approximation
for $\sim10\%$, $\sim1\%$, and $\sim0.1\%$ accuracy one has to know the
spectrum up to $\xD\sim 10^2$, $\xD\sim 10^3$, and $\xD\sim 10^4$. This shows
how important the knowledge of the solution for the photon distribution in the
distant wings is.

In the case of complete redistribution one can easily show that $\frac{\Delta
  P_{\rm S}(\nu_{\rm m})}{P_{\rm S}}=-16\% \left[\frac{\xD}{10^6}\right]^{-1}$
at $z=1100$. This implies that for $\sim 10\%$, $\sim 1\%$, and $\sim 0.1\%$
accuracy one has to know the spectrum up to $\xD\sim 10^6$, $\xD\sim 10^7$,
and $\xD\sim 10^8$. 
Let us emphasize again that these are extremely large (even unphysical) distances
from the Lyman $\alpha$ resonance.
However, it is in these regions where the value of the Sobolev escape
probability is formed.

\subsection{Effective escape probability using the time-dependent solution}
\label{sec:Pesc_def_num_Voigt_time}
With the solution \eqref{app:kin_abs_em_Sol_phys} we can also describe the
time-dependence of $\bar{n}$ within the approximation of coherent scattering
in the lab frame.
Although one does expect some modifications when accounting for partial
frequency redistribution, our computations \citep{Chlubaprep} show that the
additional correction will be dominated by the influence of
line recoil\footnote{Including atomic recoil we find a correction of $\Delta
  P/P\sim 4\%$ at $z\sim 1100$ and $\Delta P/P\sim 6\%$ at $z\sim 800$ to the
  Sobolev escape probability, which, in reasonable agreement with
  \citet{Grachev2008}, leads to $\Delta N_{\rm e}/N_{\rm e}\sim -1.2\%$ at
  $z\sim 950$.}, \changeF{which} has been addressed in \citet{Grachev2008}.
However, the time-dependent correction that is considered here turns out to be
much larger, so that we shall focus on this only.
Below we now provide a detailed discussion of the time-dependent correction in
the case of coherent scattering in the lab frame, introducing an effective
escape probability, which then can be used in computations of the cosmological
recombination history.

\subsubsection{Escape probability during recombination without redistribution
  but with full time-dependence}
\label{sec:Pesc_def_num_Voigt_time_1}
Using the time-dependent solution for the case of no redistribution,
Eq.~\eqref{app:kin_abs_em_Sol_phys}, it is possible to write
\bsub
\label{app:Nbar_kin_abs_em_Sol}
\beal
\label{app:Nbar_kin_abs_em_Sol_a}
\bar{n}^{\rm coh}(z)&=\bar{n}^{\rm pl}+\frac{c^2}{2\nu^2_{21}} \int^\infty_0\varphi(\nu) \,\Delta N^{\rm coh}_{\nu}(z)\id\nu
\nonumber
\\
&=\bar{n}^{\rm pl}+\Delta \bar{n}_{\rm em}(z)
\!\!\int^\infty_0 \!\!\varphi(\nu)\id\nu \!\!\int_{z_{\rm s}}^z \Theta^{\rm coh}(z')\,
\partial_{z'} e^{-\tau_{\rm abs}(\nu, z', z)} \id z'
\nonumber
\\
&=\bar{n}^{\rm pl}+\Delta \bar{n}_{\rm em}(z)\,(1-P^{\rm t}_{\rm em})
\\[2mm]
\label{app:Nbar_kin_abs_em_Sol_b}
P^{\rm t}_{\rm em}
&=
1+\int^\infty_0 \!\!\!\!\varphi(\nu)\id\nu \int^{z_{\rm s}}_z 
\Theta^{\rm coh}(z')
\,
\partial_{z'}e^{-\tau_{\rm abs}(\nu, z', z)}\id z',
\end{align}
\esub
where we have $\Delta \bar{n}_{\rm em}(z)=n_{\rm em}-\bar{n}^{\rm pl}$. 

Here it is very important to mention that one has to use $\Theta^{\rm coh}(z)$
as defined by Eq.~\eqref{app:kin_abs_em_Sol_phys_b} but evaluate the blackbody
distribution at the line center {\it only}, \changeF{i.e. use $\Theta^{\rm coh}(z')\approx[\tilde{N}_{\rm em}(z')-\tilde{N}^{\rm pl}_{x'_{21}}]/[\tilde{N}_{\rm em}(z)-\tilde{N}^{\rm pl}_{x_{21}}]$ with $x'_{21}=\nu_{21}/[1+z']$}.
This is necessary in order to be consistent with the formulation of line
emission and absorption processes, which, as mentioned in
Sect.~\ref{sec:abs_term}, in full equilibrium does not exactly conserve a
blackbody distribution.
A more consistent formulation will be given in a future paper, but the result
for the pure time-dependent correction should be very similar.

Eq.~\eqref{app:Nbar_kin_abs_em_Sol_a} provides the time-dependent solution for
$\bar{n}(z)$, when the ionization history is known until $z$.
However, in real calculations Eq.~\eqref{app:Nbar_kin_abs_em_Sol_a} is not
very useful, since the evaluation of the integral is rather time-consuming.
With Eq.~\eqref{app:Nbar_kin_abs_em_Sol_b} we defined an {\it effective escape
  probability}, which can be compared with the result in the full
quasi-stationary case. The differences will be due to non-stationary
contributions in the evolution of the photon distribution, and can be
iteratively used in \changeF{computations} of the recombination history.
Since the correction is expected to be small, even the first iteration should
give a rather good answer.

To obtain the difference from the Sobolev escape probability, one again has to
use the quasi-stationary solution for the 2p state, leading to relation
\eqref{eq:N_em_QS}.
With this one can eliminate $n_{\rm em}$ from \eqref{app:Nbar_kin_abs_em_Sol},
and bring the expression for $\bar{n}$ in the standard form
\eqref{eq:Nb_Seager}.
This yields
\bsub
\label{eq:in_standard_form}
\beal
\bar{n}^{\rm coh}(z)&=n_{\rm L}-P^{\rm L}_{\rm esc}\,[n_{\rm L}-\bar{n}^{\rm pl}]
\\
\label{app:P_esc_S}
P^{\rm L}_{\rm esc}&=\frac{p_{\rm d} P^{\rm t}_{\rm em}}{1-p_{\rm sc} P^{\rm t}_{\rm em}}.
\end{align}
\esub
Now $P^{\rm L}_{\rm esc}$ can be directly compared with the Sobolev escape
probability.

Looking at \eqref{app:Nbar_kin_abs_em_Sol_b} it is clear that there are two
sources for the time-dependent correction. The first comes from the
time-dependence of $\Theta^{\rm coh}$, while the \changeF{second} is due to modifications
in the absorption optical depth, $\tau_{\rm abs}$.
Below we now discuss each correction separately.

\subsubsection{Neglecting the time-dependence of $\Theta^{\rm coh}$}
\label{sec:Theta_1}
If we set $\Theta^{\rm coh}=1$ in Eq.~\eqref{app:Nbar_kin_abs_em_Sol_b} one
obtains
\beal
\label{eq:P_time_0}
P^{\rm t, 0}_{\rm em}(z)
&=\int^\infty_0 \varphi(\nu)\, e^{-\tau_{\rm abs}(\nu, z_{\rm s}, z)}\id\nu.
\end{align}
With this expression it is possible to take into account the time-dependent
corrections that are only due to the modifications of $\tau_{\rm abs}$ in
comparison to the quasi-stationary case (see Sect.~\ref{sec:ncoh}).

First, it is clear that due to the $\nu$-dependence of the absorption
cross section the total absorption optical depth depends strongly on the
initial frequency of the emitted photon.
For example, if a photon is emitted on the \changeF{blue} side of the Lyman $\alpha$
resonance, then after some redshifting it will come close to the Doppler core of
the Lyman $\alpha$ line, where it will be absorbed with extremely high
probability. Depending on the initial distance to the Doppler core, this will
take some time, during which the properties of the medium (e.g. the ionization
degree) may have changed significantly.
Similarly, photons emitted in the very distant red wing of the Lyman $\alpha$
line may redshift for a very long time, before they will be reabsorbed, if
at all.

At high $z$ the total absorption optical depth is expected to mainly \changeF{vary}
due to the changes in the number density of ionized hydrogen atoms, and at low
redshifts because of the steep drop in $p_{\rm d}$.
If for given initial frequency $\nu'$ of an emitted photon the time it takes
until this photon is reabsorbed ($\tau_{\rm abs}\sim 1$) is similar to the
Hubble time, then these changes may be important.

If the considered photon was emitted close to the Lyman $\alpha$ line center,
the absorption optical depth is dominated by its value inside the Doppler
core, where photons only travel a very short distance (a small fraction of the
Doppler width), before being reabsorbed.
In this case, the quasi-stationary approximation certainly is valid with very
high accuracy, since $\Delta \nu/\nu\ll 1$ between emission and absorption
implies $\Delta z/z\ll 1$, so that the medium has not changed very much.
However, when the photon is initially released in the distant red or blue wing
of the Lyman $\alpha$ resonance, it can redshift for a much longer time before
being reabsorbed, so that changes in the medium, in particular the
ionization degree and death probability, may play an important role.
%

\begin{figure}
\centering 
\includegraphics[width=0.9\columnwidth]
{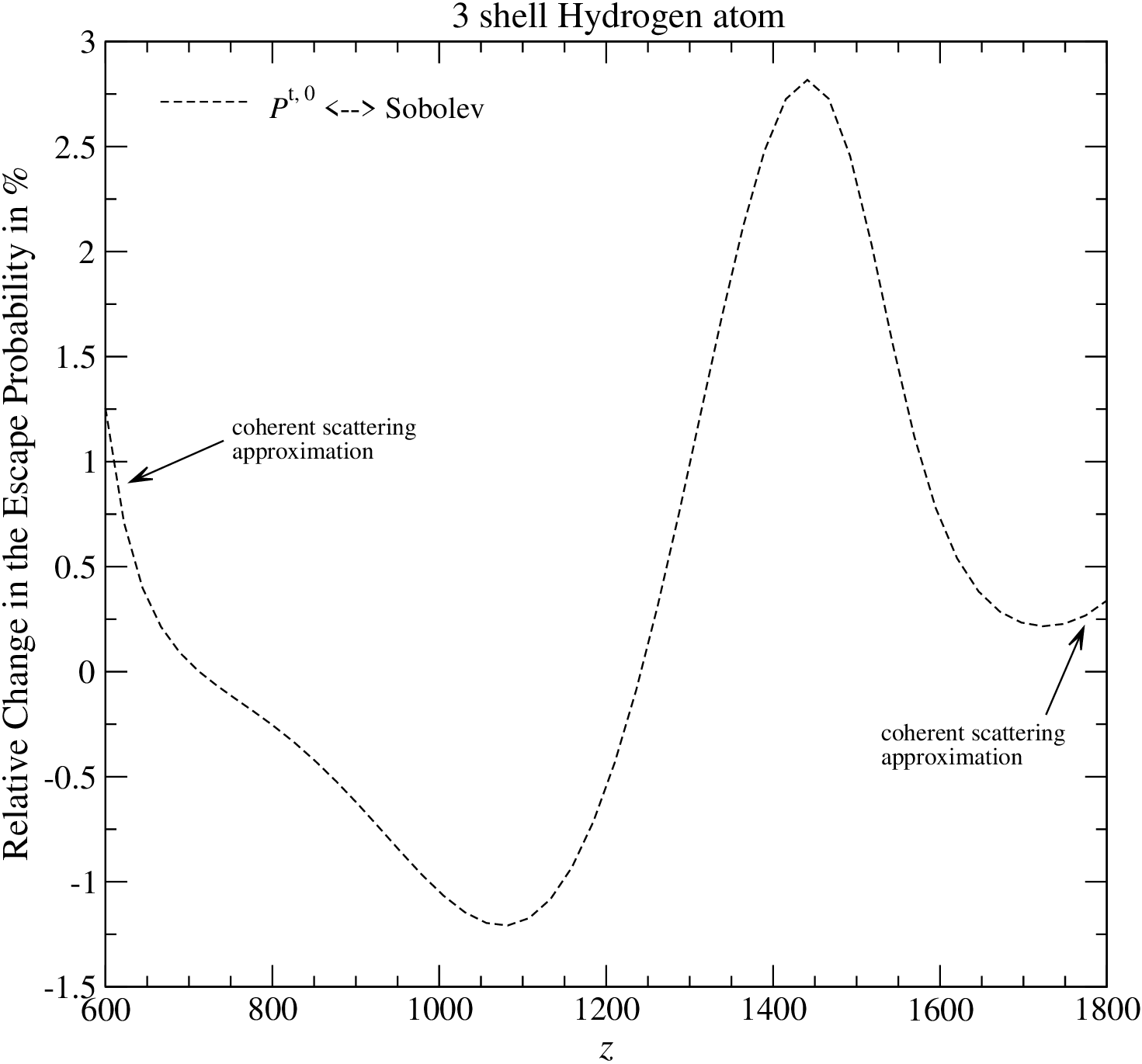}
\caption
{Differences between the escape probability $P=\frac{\pdeath P^{\rm t, 0}_{\rm
      em}}{1-\psc\,P^{\rm t, 0}_{\rm em}}$, where $P^{\rm t, 0}_{\rm em}(z)$
  is given by Eq.~\eqref{eq:P_time_0}, and the standard Sobolev escape
  probability, $P_{\rm S}$.
We used the death probability for the 3 shell hydrogen atom.
}
\label{fig:DP_P.time.P0}
\end{figure}
In Fig.~\ref{fig:DP_P.time.P0} we show the direct comparison of the escape
probability that follows from Eq.~\eqref{eq:P_time_0} with the Sobolev escape
probability.
At very low and very high redshifts the correction due to the pure
time-dependence of $\tau_{\rm abs}$ becomes very small. The difference that is
seen close to $z\sim 600$ and $z\sim 1800$ is only related to the correction
coming from the coherent scattering approximation (see
Fig.~\ref{fig:DP.P_d.P_S}).
In both cases this behavior can be explained by the fact that the importance
of the wings for the total value of the escape probability decreases. Photons
escape directly from the Doppler core, so that the contributions to the value
of $P^{\rm t, 0}_{\rm em}(z)$ can be considered quasi-stationary.

To understand the behavior at intermediate redshift, it is important that
before the maximum of $\tau_{\rm d}$ around $z_{\rm max}\sim 1300$
(cf. Fig.~\ref{fig:tau_d_S}), one expects that independent\footnote{This
  statement is not completely correct, since in Eq.~\eqref{app:tau_abs_nut} we
  do take into account the factor $1/\nu$. However, this only affects the very
  distant blue wing.}
of the considered frequency, $\tau_{\rm abs}(\nu, z', z)$ is smaller than in
the quasi-stationary approximation $\tau^{\rm qs}_{\rm abs}(\nu, z',
z)=\tau_{\rm d}(z)\,\int^{\nu'}_\nu\varphi(\nu'')\id\nu''$.
This is simply because at $z'>z\gtrsim z_{\rm max}$ the value of $\tau_{\rm
  d}(z')\lesssim\tau_{\rm d}(z)$.
At those times the time-dependent modifications of $\tau_{\rm abs}$ should
therefore result in a positive correction to the effective escape probability
(cf. Fig.~\ref{fig:DP_P.time.P0}).
With a similar argument, at redshift $z\lesssim 1300$ the correction in the
escape probability should be negative, as is seen in
Fig.~\ref{fig:DP_P.time.P0}.
%

\begin{figure*}
\centering 
\includegraphics[width=0.9\columnwidth]{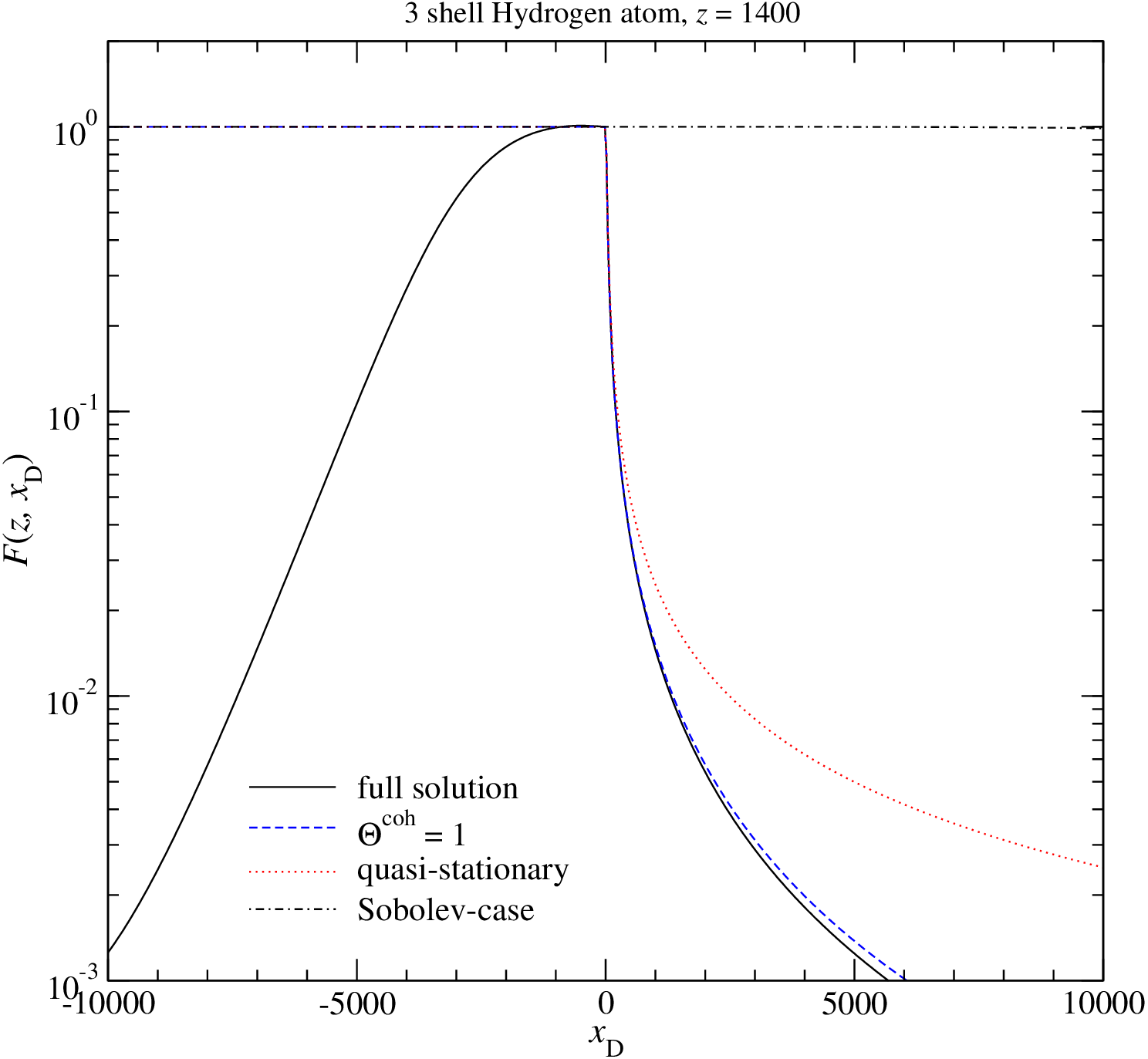}
\includegraphics[width=0.9\columnwidth]{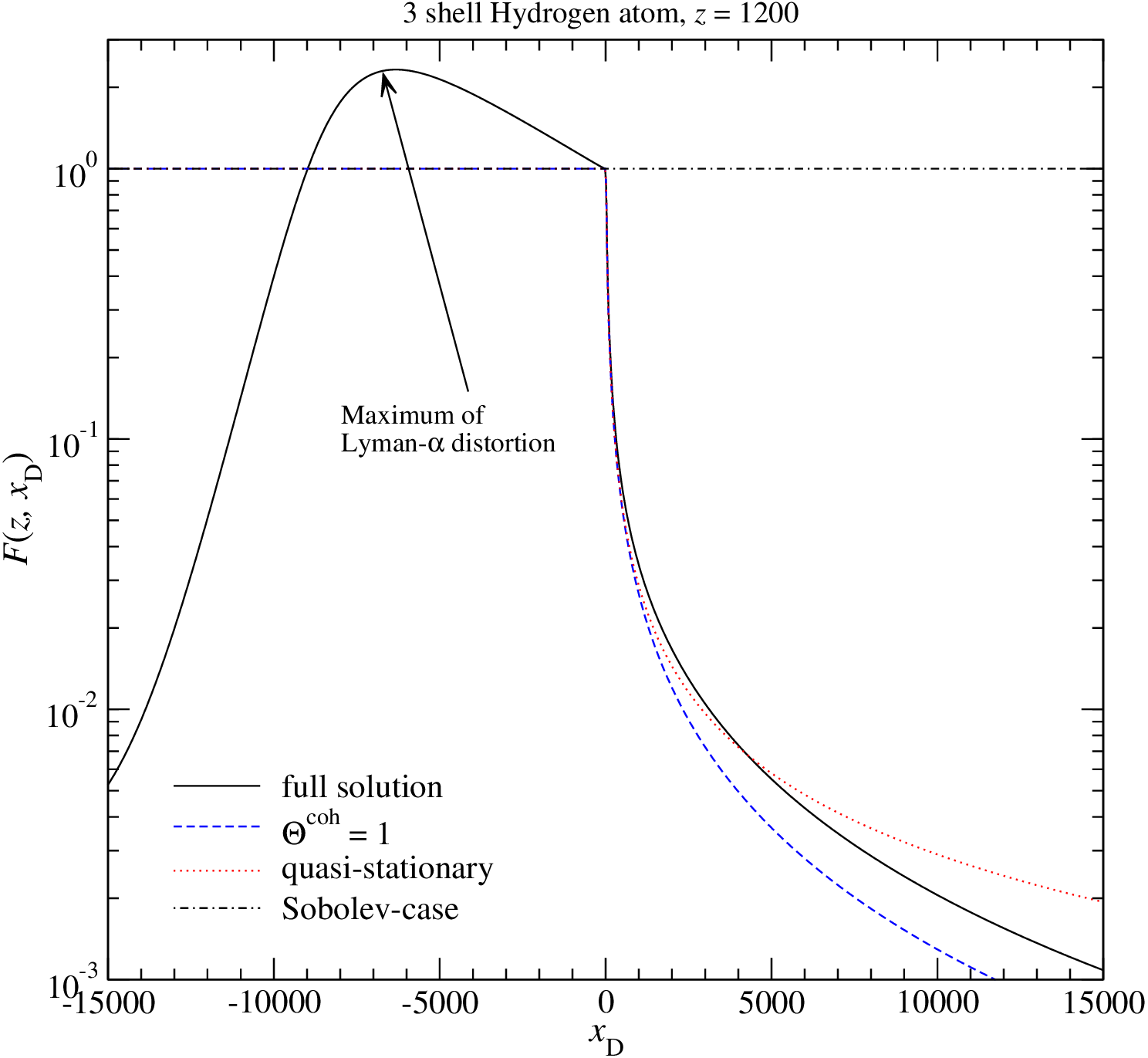}
\\
\includegraphics[width=0.9\columnwidth]{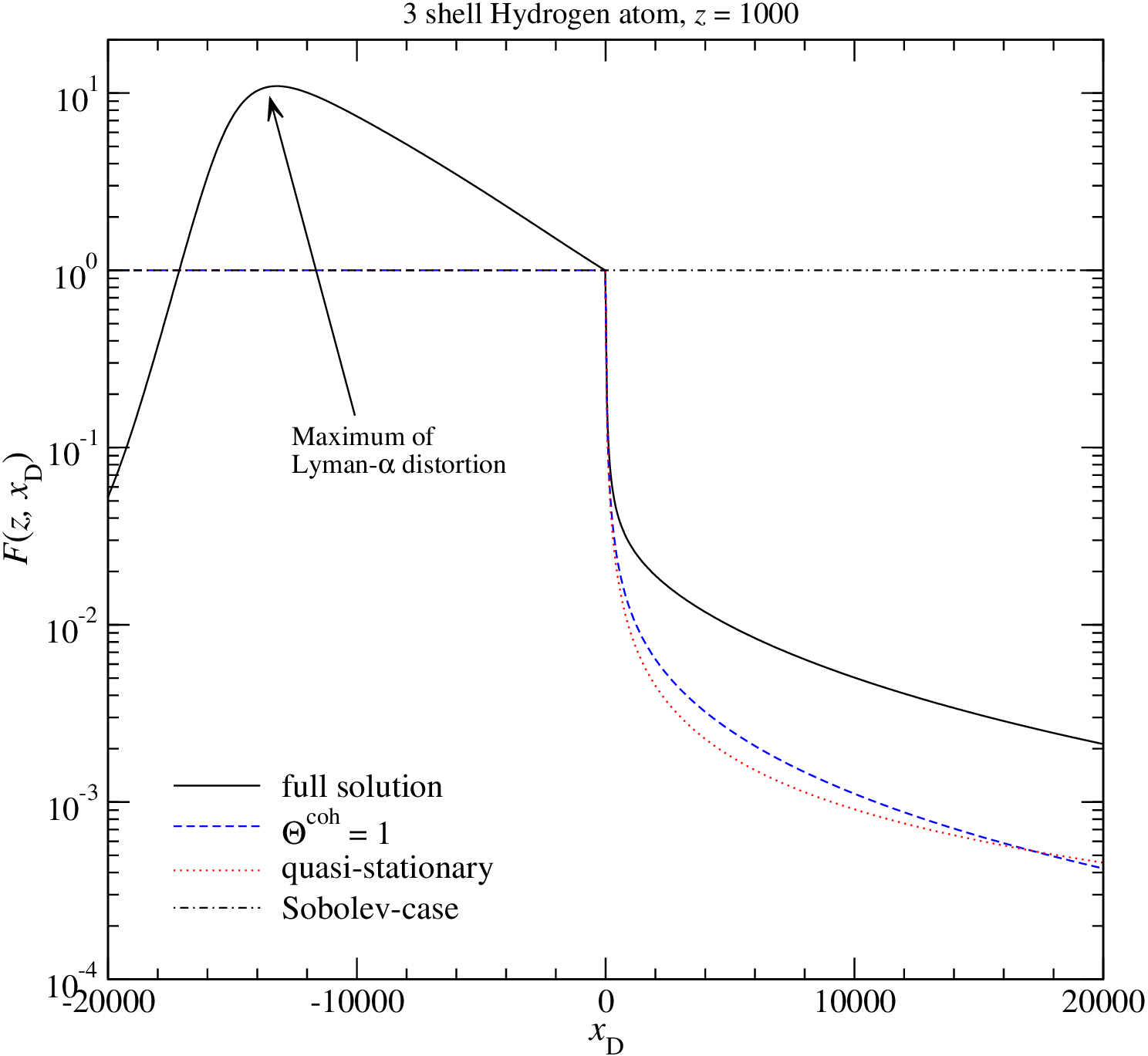}
\includegraphics[width=0.9\columnwidth]{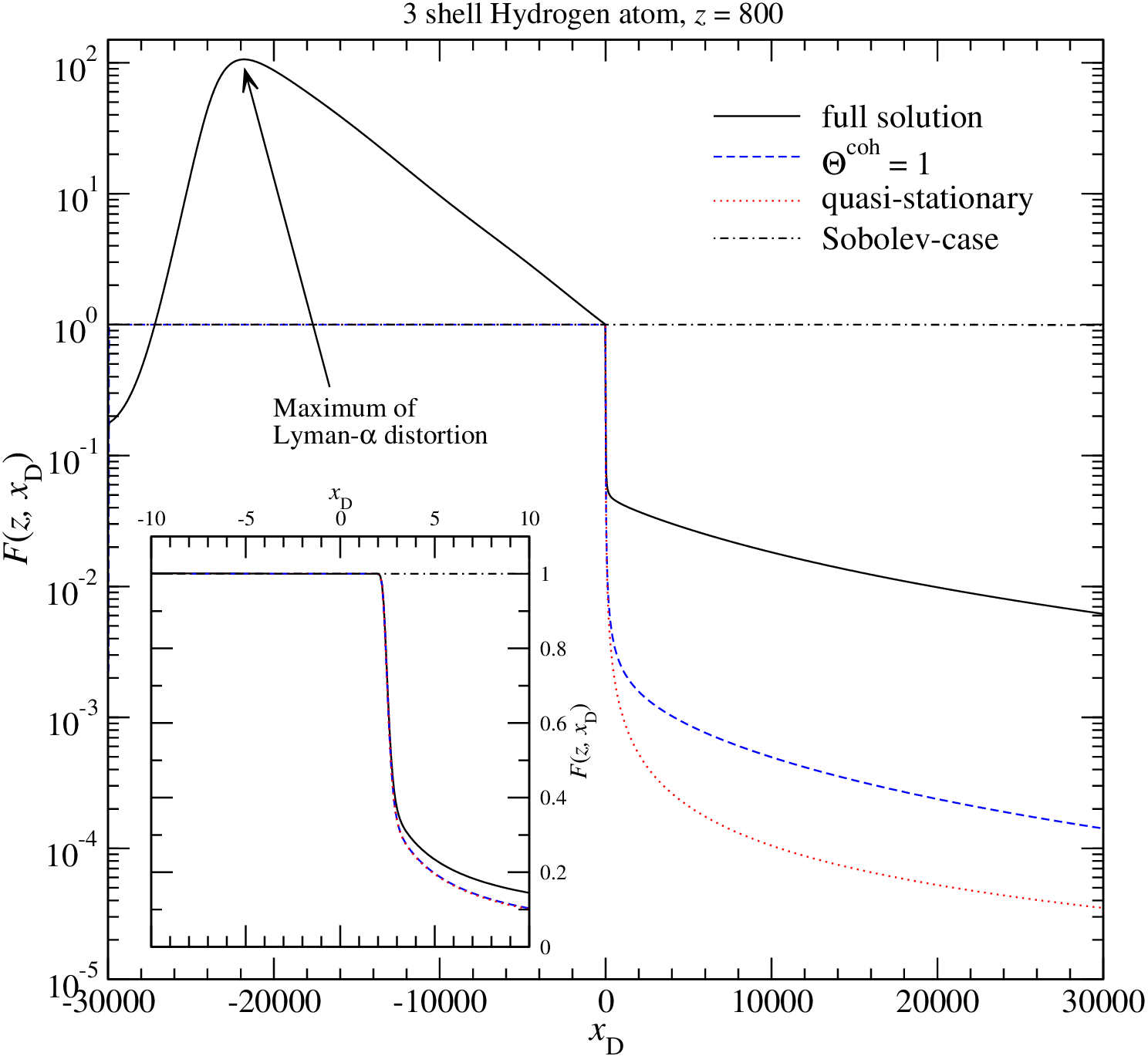}
\caption
{ The spectral distortion $\Delta N_\nu$ far away from the Lyman $\alpha$ line
  center at different stages of hydrogen recombination. In all cases
  redistribution of photons over frequency was neglected. We normalized the
  distortion to unity at the Lyman $\alpha$ frequency, i.e. $F\equiv\Delta
  N_\nu(z)/\Delta N_{\nu_{21}}(z)$.
The dotted curves give the spectral distortion in the quasi-stationary
approximation, $F^{\rm qs}(\nu)=1-e^{-\tau_{\rm d}}e^{\tau_{\rm d}\chi(\nu)}$.
For the dashed curves we neglected the variation of $\Theta^{\rm coh}$ in the
solution \eqref{app:kin_abs_em_Sol_phys}, leading to $F_0(\nu)=1-e^{-\tau_{\rm
    abs}(\nu, z_{\rm s}, z)}$, with $z_{\rm s}\rightarrow \infty$.
%
%
The solid lines show the spectral distortion in the time-dependent case,
i.e. $F(\nu)=F_0(\nu)+\Delta F(\nu)$, where $\Delta F(\nu)$ is defined by
Eq.~\eqref{app:DP_time_b}.
For comparison, the dash-dotted curves give the spectral distortion in the
Sobolev approximation, $F^{\rm S}(\nu)=1-e^{-\tau_{\rm S}}e^{\tau_{\rm
    S}\chi(\nu)}$.
We also indicated the position of the maximum of the redshifted Lyman $\alpha$
distortion, which appeared at redshift $z\sim 1400$.
}
\label{fig:F_time.z}
\end{figure*}

\begin{figure*}
\centering 
\includegraphics[width=0.9\columnwidth]{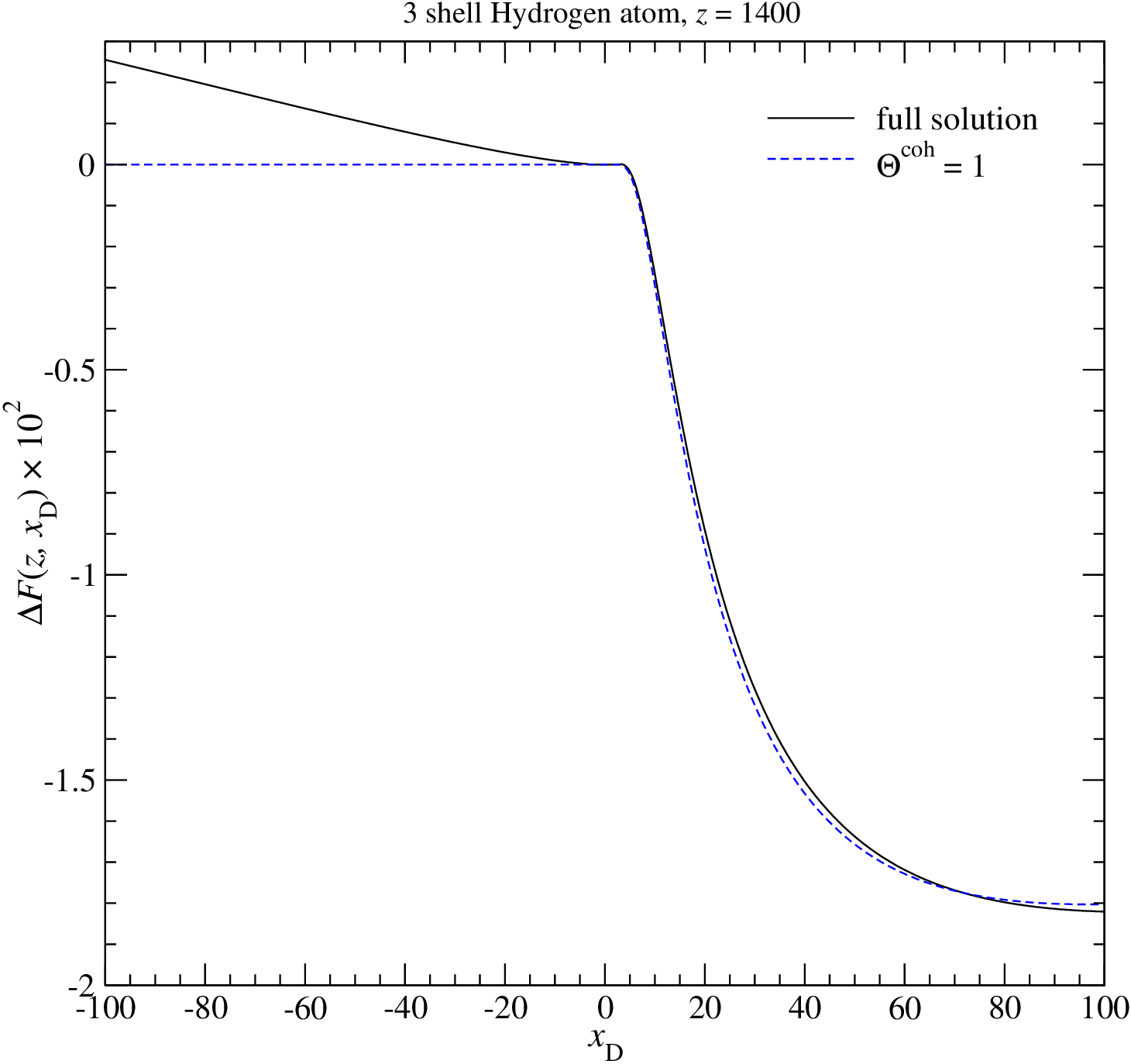}
\includegraphics[width=0.9\columnwidth]{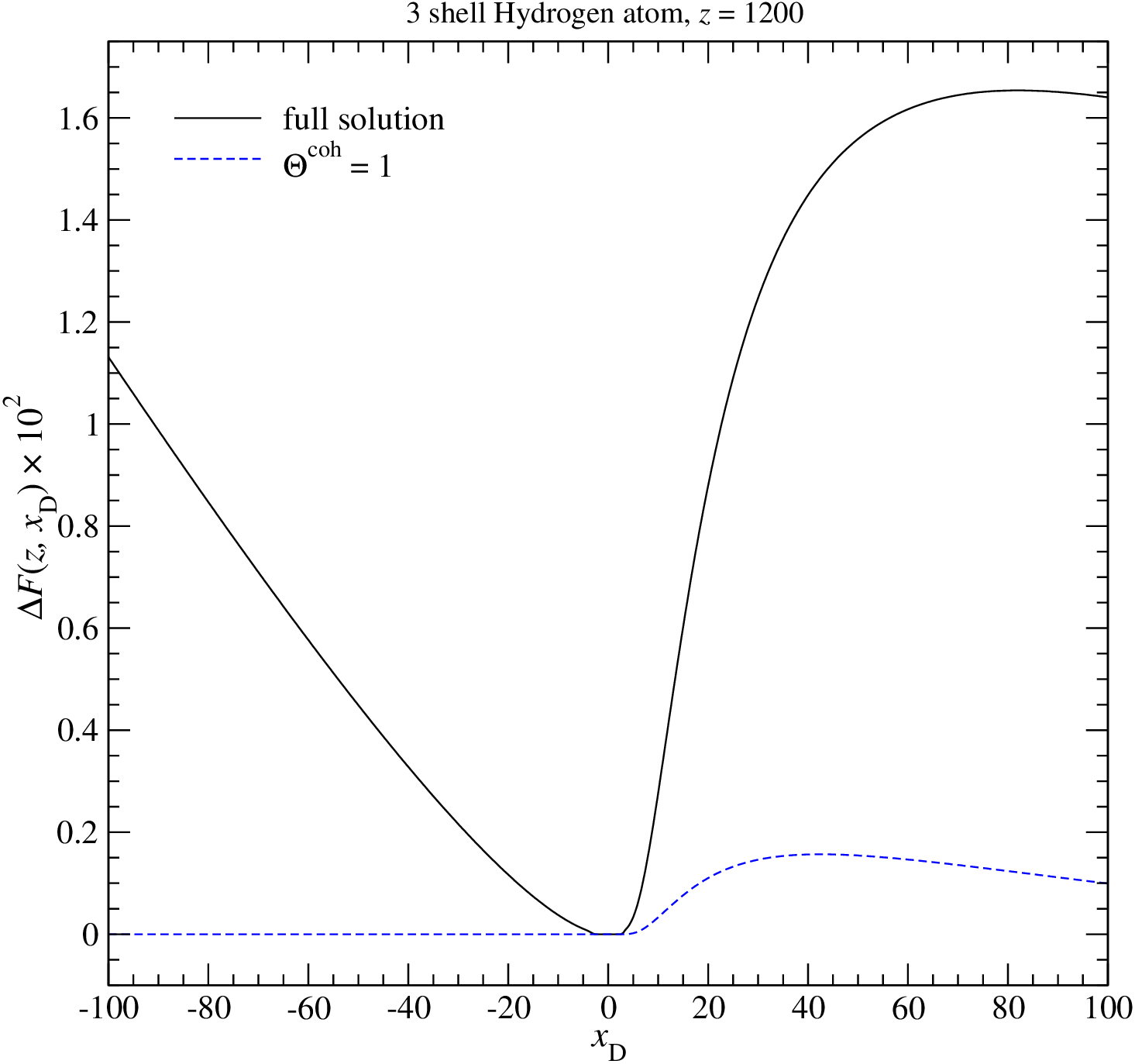}
\\
\includegraphics[width=0.9\columnwidth]{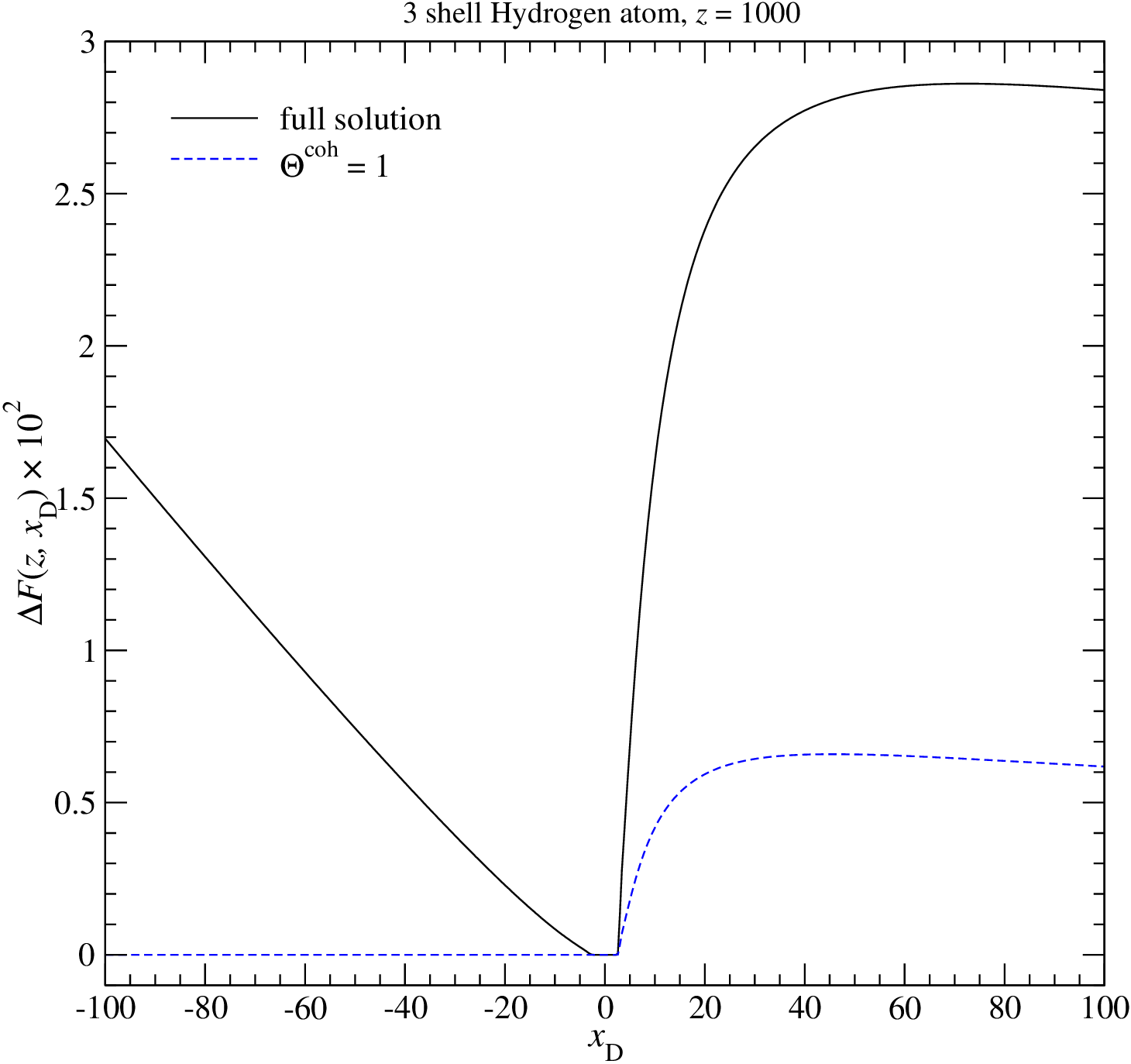}
\includegraphics[width=0.9\columnwidth]{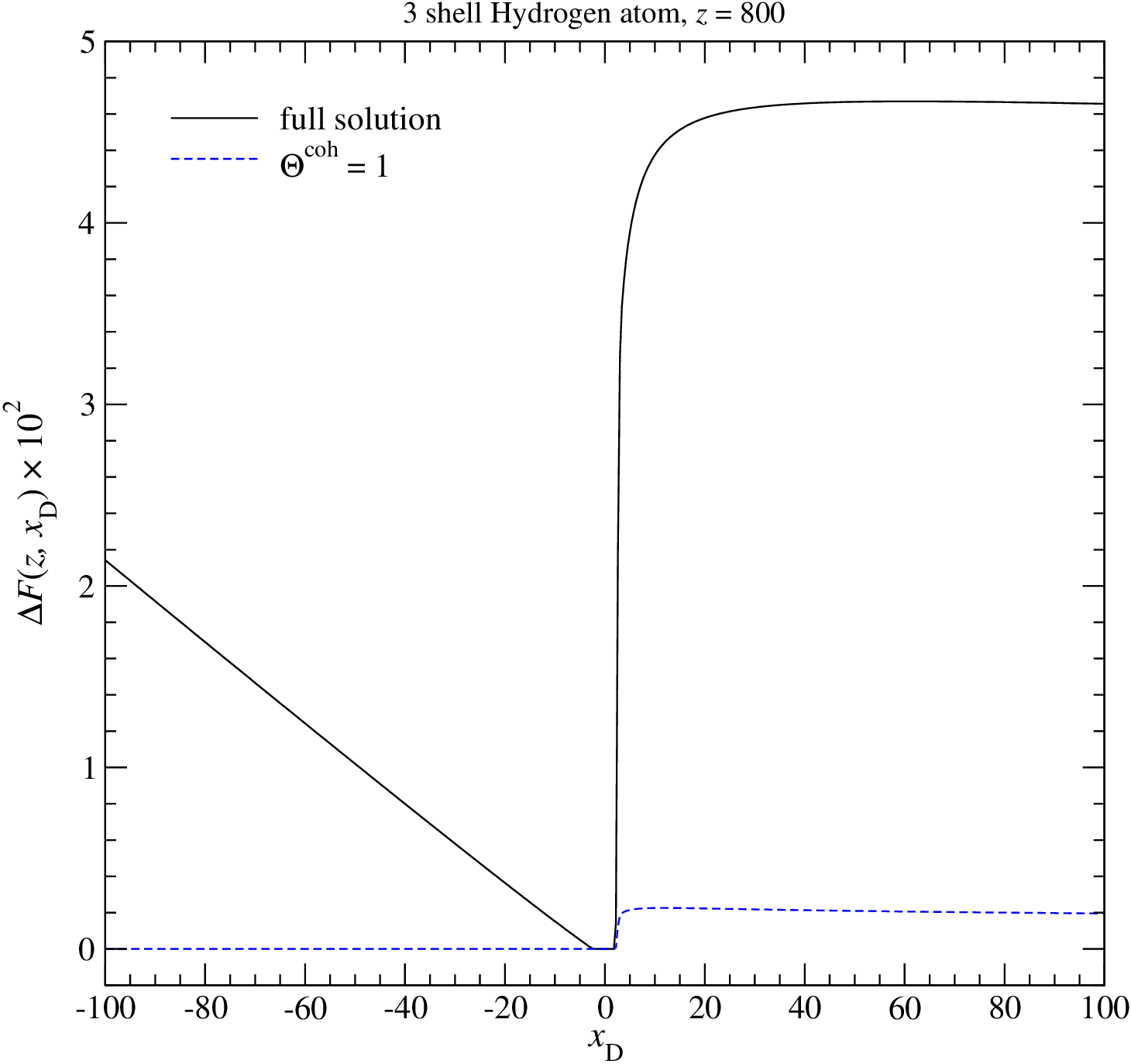}
\caption
{ The spectral distortion $\Delta N_\nu$ close to the Lyman $\alpha$ line
  center at different stages of hydrogen recombination. In all cases
  redistribution of photons over frequency was neglected. We normalized the
  distortion to unity at the Lyman $\alpha$ frequency, i.e. $F\equiv\Delta
  N_\nu(z)/\Delta N_{\nu_{21}}(z)$ and show the difference to the
  quasi-stationary solution, $F^{\rm qs}(\nu)=1-e^{-\tau_{\rm d}}e^{\tau_{\rm d}\chi(\nu)}$.
For the dashed curves we neglected the variation of $\Theta^{\rm coh}$ in the
computation of the escape probability, while for the solid lines the spectral
distortion in the time-dependent case was taken into account.
}
\label{fig:DF_time.z}
\end{figure*}

\subsubsection{Correction due to the time-dependence of $\Theta^{\rm coh}$}
\label{sec:Theta_corr}
\changeF{In Sect.~\ref{sec:Theta_1} we have neglected the time-dependence of $\Theta^{\rm coh}$. This factor describes how much the photon emission process varies as a function of time, which in the present approximation is independent of frequency (see comment in Sect.~\ref{sec:Pesc_def_num_Voigt_time_1}).}

With Eq.~\eqref{eq:P_time_0} and \eqref{app:Nbar_kin_abs_em_Sol_b} one can
define
\bsub
\label{app:DP_time}
\beal
\label{app:DP_time_a}
\Delta P^{\rm t}_{\rm em}
&=
\int^\infty_0 \varphi(\nu)\id\nu \int^{z_{\rm s}}_z 
\left[\Theta^{\rm coh}(z')-1\right]
\,
\partial_{z'}e^{-\tau_{\rm abs}(\nu, z', z)}\id z'
\nonumber
\\
&=-\int^\infty_0 \!\!\!\!\varphi(\nu)\,\Delta F(\nu)\id\nu
\\[2mm]
\label{app:DP_time_b}
\Delta F(\nu)
&=\int_{z_{\rm s}}^z 
\left[\Theta^{\rm coh}(z')-1\right]
\,
\partial_{z'}e^{-\tau_{\rm abs}(\nu, z', z)}\id z'.
\end{align}
\esub
With this expression it is now possible to calculate\footnote{The evaluation
  of the integral \eqref{app:DP_time_a} is rather cumbersome. It is most
  important that for a fixed frequency $\nu$ at redshift $z$ the inner
  integral varies most strongly at $z'\sim \max(z, \nu_{21}(1+z)/\nu)$.}
the additional correction to the escape probability coming from the variation
of $\Theta^{\rm coh}$ over time.
%

\label{sec:F_time}
In order to understand the final result we first consider the behavior of the
inner integrand \eqref{app:DP_time_a} at different stages of hydrogen
recombination.
\changeF{Since} the function 
$F(\nu)=\int_{z_{\rm s}}^z \Theta^{\rm coh}(z')\,\partial_{z'}e^{-\tau_{\rm
    abs}(\nu, z', z)}\id z'$ is identical to the Lyman $\alpha$ spectral
distortion in the no redistribution approximation, but normalized to its value
at the line center, i.e. $F(\nu)\equiv \Delta N_\nu(z)/\Delta
N_{\nu_{21}}(z)$, it is illustrative to define
\bsub
\label{app:F_time}
\beal
\label{app:F_time_a}
F^{\rm qs}(\nu)&=1-e^{-\tau_{\rm d}}e^{\tau_{\rm d}\chi(\nu)}
\\
\label{app:F_time_b}
F_0(\nu)&=1-e^{-\tau_{\rm abs}(\nu, z_{\rm s}, z)}
\\
\label{app:F_time_c}
F(\nu)&=F_0(\nu)+\Delta F(\nu),
\end{align}
\esub
in addition to Eq.~\eqref{app:DP_time_b}. Here $F^{\rm qs}(\nu)$ represents
the Lyman $\alpha$ spectral distortion in the full quasi-stationary
approximation, for $F_0(\nu)$ only the variation of $\Theta^{\rm coh}$ is
neglected, and $F(\nu)$ is the time-dependent distortion in the no
redistribution approximation.
From these functions the corresponding escape probabilities can be obtained by
$P_{\rm esc}=1-\int\varphi(\nu) F(\nu)\id \nu$, which then can be inserted in
Eq.~\eqref{app:P_esc_S} when comparing with $P_{\rm S}$.
Therefore it is clear that if $F(\nu)<F^{\rm qs}(\nu)$ at all frequencies, the
resulting effective escape probability should be slightly larger than $P_{\rm
  d}$.

In Fig.~\ref{fig:F_time.z} and \ref{fig:DF_time.z} we illustrate the behavior
of the functions \eqref{app:F_time} at different stages of hydrogen
recombination.
We used the solution for the populations in the 3 shell case as given by our
multi-level code \citep{Jose2006, Chluba2007}.
As expected, in all cases $F^{\rm qs}(\nu)$ and $F_{0}(\nu)$ are very close to
unity at $\xD\lesssim 0$ and then drops very fast toward zero at $\xD\gtrsim
0$.
Also Fig.~\ref{fig:DF_time.z} clearly shows that $F^{\rm qs}(\nu)\approx F_{0}(\nu)$
in the red wing and the Doppler core of the Lyman $\alpha$ resonance.
This is expected, since at $\nu\lesssim \nu_{21}$ always $\tau_{\rm abs}\gg
1$, so that its exact value does not matter.
Furthermore we can observe a change in the sign of the difference
$F_{0}(\nu)-F^{\rm qs}(\nu)$ in the blue wing when going from high to lower
redshift.
At $z=1400$ one can clearly see that $F_{0}(\nu)<F^{\rm qs}(\nu)$ in the range
$0\lesssim \xD\lesssim 100$, so that $P^{\rm t, 0}_{\rm em}(z)>P_{\rm d}$ is
expected, in agreement with the results presented in
Fig.~\ref{fig:DP_P.time.P0}.
On the other hand, in all the other cases shown $F_{0}(\nu)>F^{\rm qs}(\nu)$
at $0\lesssim \xD\lesssim 100$ so that one should find $P^{\rm t, 0}_{\rm
  em}(z)<P_{\rm d}$, again confirming the results given in
Fig.~\ref{fig:DP_P.time.P0}.

Because of the steep drop of $F^{\rm qs}(\nu)$ and
$F_{0}(\nu)$ at a few Doppler width above the line center, the main
contribution to the escape probability clearly comes from rather close to the line
center.
However, at the level of percent the shape of the distortion up to a few
hundred or thousand Doppler widths is important.

If we now look at the spectral distortion in the time-dependent approximation,
$F(\nu)=F_{0}(\nu)+\Delta F(\nu)$, we can see that at {\it all} stages the
variations of $\Theta^{\rm coh}$ with redshift become important outside the
Doppler core.
From Fig.~\ref{fig:F_time.z} we can distinguish in more detail the following regimes:
(i) at redshifts $z\gtrsim 1400$ the distant wing distortion is {\it smaller}
than in the quasi-stationary approximation. This is because at redshifts much
before the time under consideration the emission in the Lyman $\alpha$
transition was very inefficient, so that until then not many photons can have
appeared or reached large distances from the Lyman $\alpha$ line. The slope of
the red wing distortion is positive close to the line center;
(ii) at redshifts $z\lesssim 1400$ the distortion in the blue wing and nearby
red wing is {\it greater} than in the quasi-stationary approximation.
The production rate of Lyman $\alpha$ photons has already passed its maximum
(at $z\sim 1400$), so that at the current line center {\it fewer} photons are
produced the lower the redshift becomes. The slope of the red wing distortion
is negative close to the line center.

\begin{figure}
\centering 
\includegraphics[width=0.9\columnwidth]
{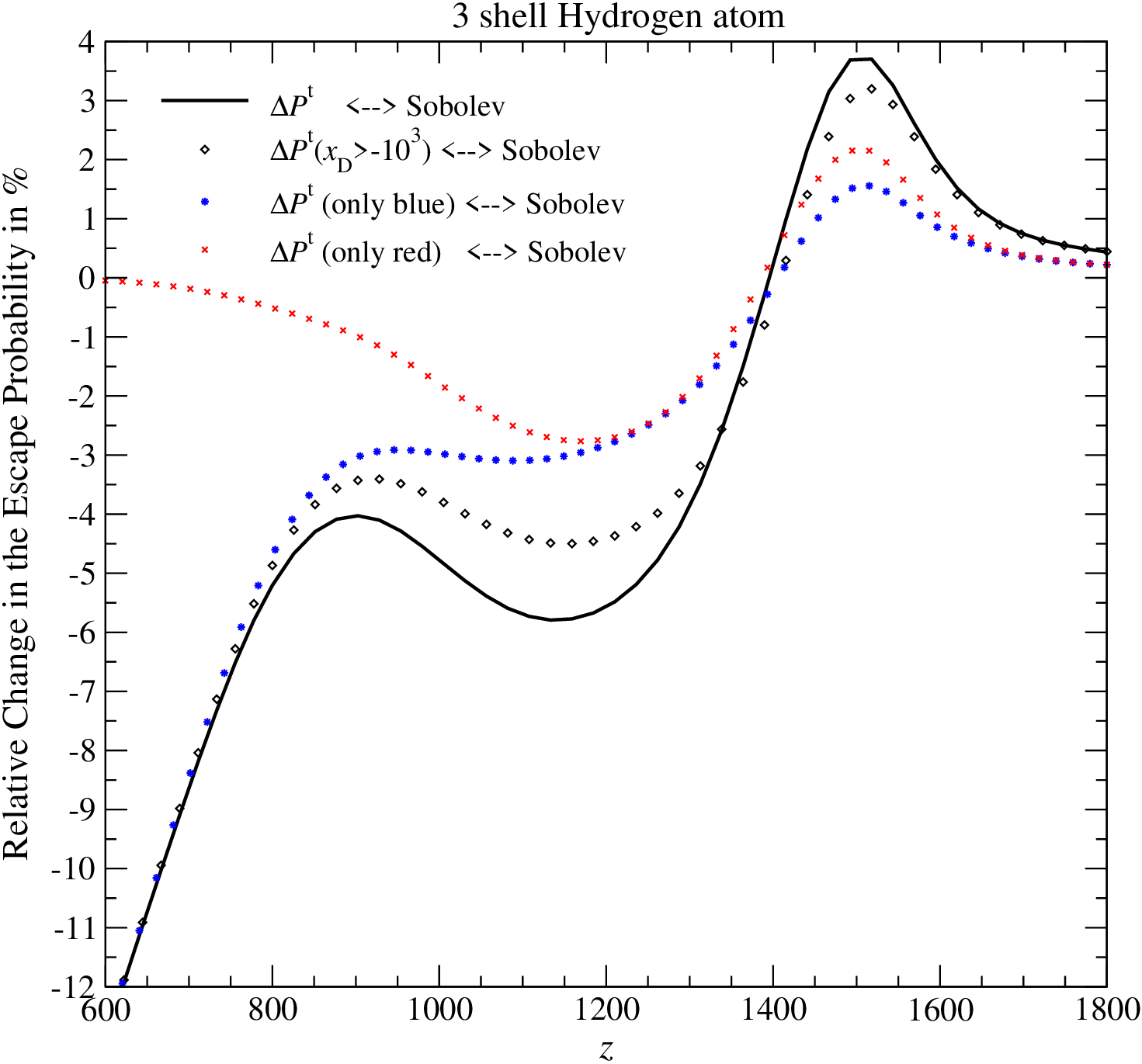}
\caption
{Correction to the Sobolev escape probability due to variations of
  $\Theta^{\rm coh}$. For the solid curve we used $\Delta P^{\rm t}_{\rm
    em}(z)$ according to Eq.~\eqref{app:DP_time}.
  The boxed curve was computed with Eq.~\eqref{app:DP_time}, but including the
  variation of $\Theta^{\rm coh}$ only at $\xD\geq -10^3$.
The other two curves were obtained setting $\Theta^{\rm coh}=1$ in the red
wing (stars) or blue wing (crosses), respectively.
}
\label{fig:DP_P.time}
\end{figure}

It is also clear that in case (i) the value of $\Delta\bar{n}$ is {\it smaller}
than in the quasi-stationary approximation, while it is expected to be {\it larger}
in case (ii). According to the definition~\eqref{app:Nbar_kin_abs_em_Sol} this
implies that in the former case the effective escape probability is {\it higher}
than in the quasi-stationary approximation, while it is {\it lower} in the latter
case.
In Fig.~\ref{fig:DP_P.time} we can see that these expectations are true (see
solid line). The total correction due to excess or a deficit of photons leads to
a total decrease of the effective escape probability at $z \lesssim 1400$
that reaches $\Delta P/P\sim -5.8\%$ at $z\sim 1140$, while it results in an
increase of $\Delta P/P\sim +3.7\%$ at $z\sim 1510$.

Although in the escape integral the distortion in the vicinity of the
line center mainly contributes, at the percent level the distant wings are also
important. As we have seen in Fig.~\ref{fig:F_time.z} the red wing distortion
due to the Lyman $\alpha$ transition can exceed the distortion close to the
line center by a large amount. In this case the question is how much the very
distant wings actually contribute to the total correction shown in
Fig.~\ref{fig:DP_P.time}.
For this we computed $\Delta P^{\rm t}_{\rm em}(z)$, but excluding the
correction at $\xD\leq -10^3$. Looking at the boxed curve in
Fig.~\ref{fig:DP_P.time} shows that the very distant red wings
contribute about $\Delta P/P\sim 0.6\%$ at $z\sim 1500$, and $\Delta P/P\sim
-1.3\%$ at $z\sim 1100$.
This is an important point, since in the very distant wings other
processes related to the formulation of the problem will also become important
(i.e. due to changes in the absorption profiles, when considering the problem
as a two-photon process), so that one expects additional revisions for
contributions from the very distant wings.
However, here the corrections mainly seems to come from regions \changeF{in the vicinity ($|\xD|\lesssim 10^2-10^3$) of} the
Lyman $\alpha$ line center.

%
At low redshifts ($z\lesssim 800-900$) one can observe an additional strong
decrease in the effective escape probability. This is due to the additional
re-excitation of electrons by the distortion on the blue side of the
Lyman $\alpha$ resonance.
Most of the photons in this part of the spectrum have been emitted much
earlier, at times around the maximum of the Lyman $\alpha$ emission ($z\sim
1400$).
This also explains the huge difference to the quasi-stationary \changeF{solution:
as} one can see in Fig.~\ref{fig:F_time.z}, at $z\lesssim 800$ the amount of
photons exceeds the spectral distortion obtained in the quasi-stationary
approximation by about two orders of magnitude. The spectral distortion is
only a factor of $\sim 100$ below the emission in the line center.
Looking at Fig.~\ref{fig:DF_time.z}, very close to the Lyman $\alpha$
line center some differences also are visible, which at the percent level do
matter.

To show that the distortions on the blue side of the Lyman $\alpha$ are
responsible for this re-excitation we also computed the correction only
including the non-stationary contributions for the red side, but setting
$\Theta^{\rm coh}=1$ for evaluations on the blue side.
The result is also shown in Fig.~\ref{fig:F_time.z} (stars). 
For completeness we also gave the curve when {\it only} including the
corrections on the blue side of the line.
As one can see, at $z\gtrsim 1100$ the red and blue wing corrections are very
similar. However, at low redshifts the blue wing correction clearly dominates,
supporting the statement made above.
Again one can expect some changes in the conclusions when treating the problem
in the full two-photon formulation, since the emission of photons at high
frequencies will be significantly less in the two-photon treatment, simply
due to the fact that due to energy conservation the emission profiles do not
extend to arbitrarily high frequencies \citep[e.g. see][]{Chluba2008a}.
However, the corrections to the escape probability at $z\lesssim 800$ do not
propagate very strongly to the ionization history, so that here we do not
consider this unphysical aspect of the solution any further.

\begin{figure}
\centering 
\includegraphics[width=0.9\columnwidth]
{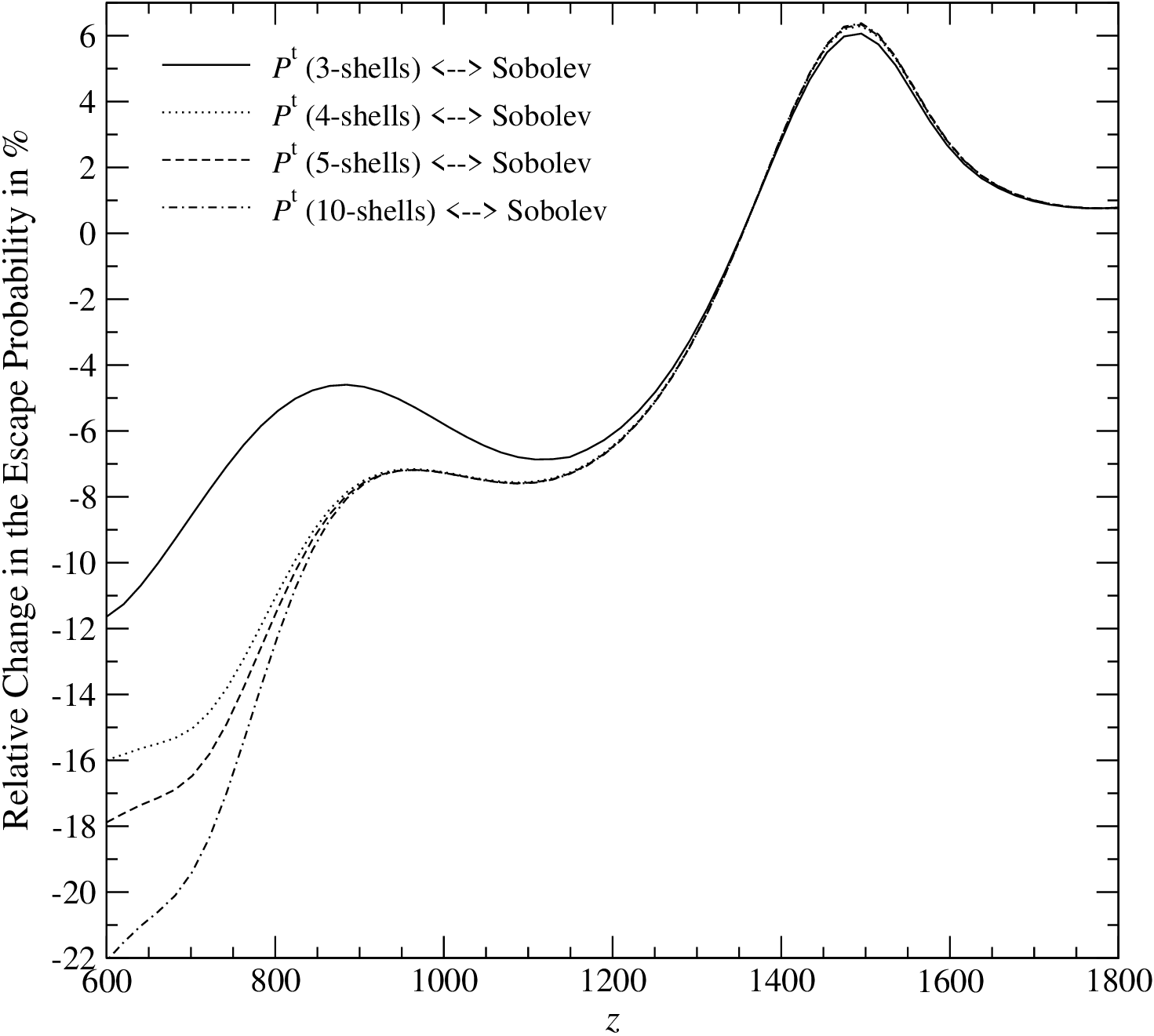}
\caption
{Total correction to the Sobolev escape probability. All curves were computed
  using the time-dependent solution according to
  Eq.~\eqref{app:Nbar_kin_abs_em_Sol_b}, but including a different number of
  shells.
}
\label{fig:DP_P.time.nS}
\end{figure}
\subsubsection{Dependence of the effective escape probability on the included
  number of hydrogen shells}
\label{sec:P_nS}
Although the death probability $\pdeath$ does not dependent on the solution of
the recombination problem, the amount of fresh electrons injected into
the Lyman $\alpha$ line depends on the populations of the excited states.
Therefore the strength of the Lyman $\alpha$ line strongly depends on the
total number of shells that are included \citep{Jose2006}.  Here in particular
the low-redshift tail ($z\lesssim 800$) will be affected, and hence there one
also expects changes in the correction to the effective escape probability.

In Fig.~\ref{fig:DP_P.time} we show the differences in the escape probability
when including more shells. We used the numerical solution for the excited
levels as obtained with our multi-level hydrogen code \citep{Chluba2007}.
At redshifts $z\gtrsim 1200$ the result is practically unaffected by the total
number of hydrogen shells that are included. In particular the result seems
converged when including $\sim 4-5$ shells, leading to a total corrections of
$\Delta P/P\sim -7.6\%$ at $z\sim 1100$, while it results in an increase of
$\Delta P/P\sim +6.4\%$ at $z\sim 1490$.
At redshifts $900\lesssim z \lesssim 1200$, some small changes are still visible when
including more than 3 shells, but again the result seems to remain unchanged
when including more than $\sim 4-5$ shells.
For the computation of the CMB power spectra the corrections in this range are
most important (see Sect. \ref{sec:changes_in_Xe}).

At lower redshifts, however, the result still changes notably. At $z\sim 600$
the correction increases by about $6.4\%$ when including 5 shells, and for 10
shells even by about $10\%$. This can be explained when realizing that the
total emission in the Lyman $\alpha$ line at low redshifts becomes less
when including more shells. Therefore the wing emission from redshifts around
the maximum of the Lyman $\alpha$ emission ($z\sim 1400$), which practically
remains unchanged, becomes more important, being able to re-excite the
2p state as explained in the previous paragraph.
As we mentioned already this aspect will probably be affected when including
corrections to the emission and absorption profiles according to the
two-photon formulation.
Furthermore, as we will see in the next Sect. this low redshift tail is not so
important for the predictions of the CMB power spectra.

\begin{figure}
\centering 
\includegraphics[width=0.9\columnwidth]
{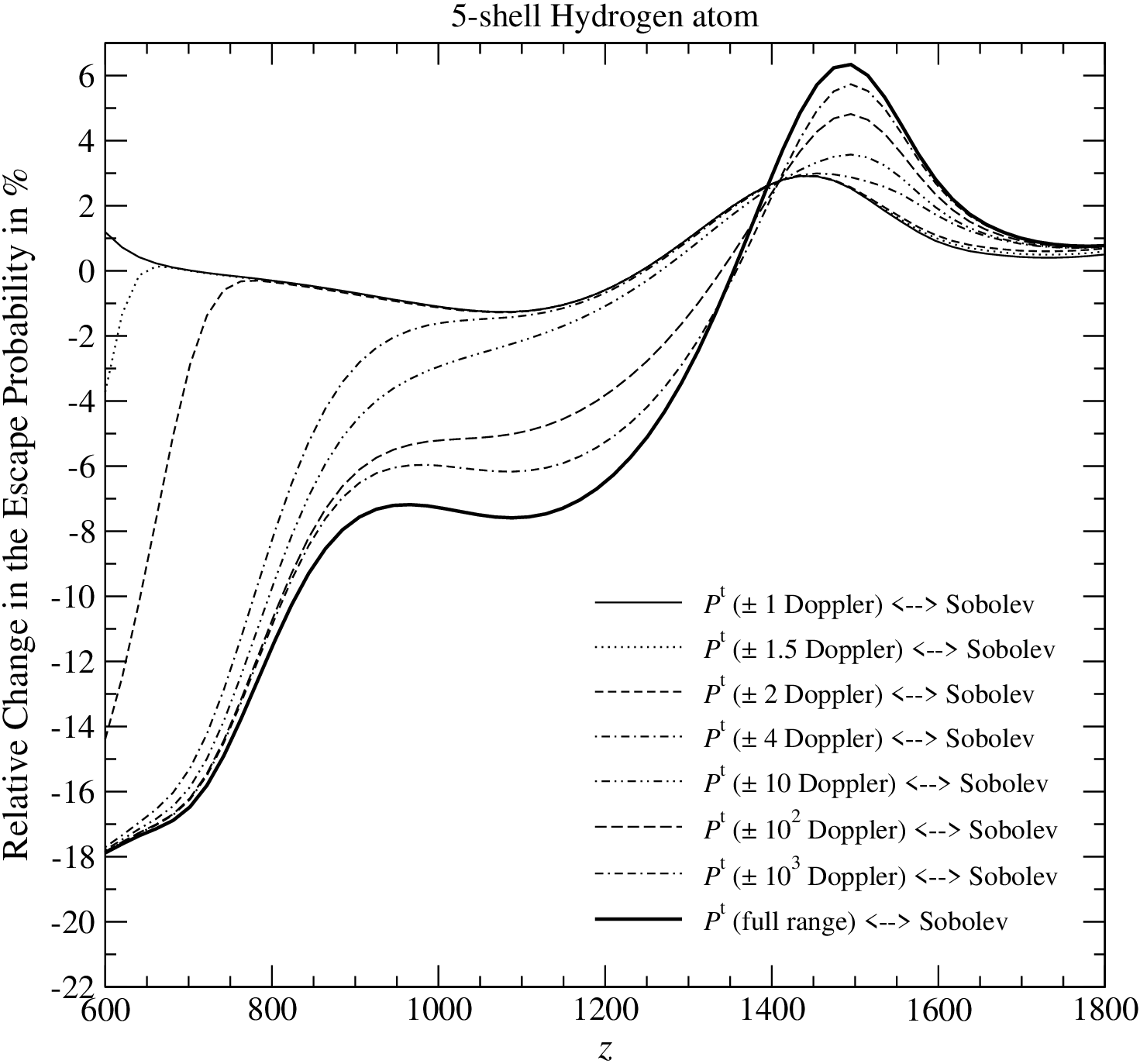}
\caption
{\change{Total correction to the Sobolev escape probability for the 5 shell
    hydrogen atom. All curves were computed using the time-dependent solution,
    $P^{\rm t}_{\rm em}=P^{\rm t, 0}_{\rm em}+\Delta P^{\rm t}_{\rm em}$
    according to Eqs.~\eqref{app:Nbar_kin_abs_em_Sol_b}, \eqref{eq:P_time_0}
    and \eqref{app:DP_time}. The term $P^{\rm t, 0}_{\rm em}$ was fully
    included, but for the contribution from $\Delta P^{\rm t}_{\rm em}$ the
    time-dependent correction was only taken into account for a given central
    region $|\xD|\leq \rm const$ around the resonance, as labeled
    respectively.}
}
\label{fig:DP_P.part.nS_5}
\end{figure}
\subsubsection{Dependence of the effective escape probability on the distance
  from the line center}
\label{sec:P_wings}
\change{
As a last point, we want to answer the question of where in the case of $\Delta
P^{\rm t}_{\rm em}$ the main correction actually comes from. We have
already seen in Sect.~\ref{sec:F_time} that the very distant red wing
contributes at the level of percent.
Also we have seen that at redshifts $z\gtrsim 800-900$ the blue and red sides
of the Lyman $\alpha$ line give similar contributions, while at low redshift
due to the self-feedback the blue wing clearly dominates.

In Fig.~\ref{fig:DP_P.part.nS_5} we show the total correction to the
Sobolev escape probability when only including the time-dependent correction
for a given central region around the resonance in the computation of $\Delta
P^{\rm t}_{\rm em}$.
It is obvious that the innermost Doppler core ($\pm 1$ Doppler width) does not
contribute much to the result. This is expected, since there the photon
distribution should evolve as in the quasi-stationary approximation, with
very high accuracy.
This fact can also be seen in Fig.~\ref{fig:DF_time.z}, where close to the
line center the deviation of the photon distribution for the quasi-stationary
solution is very small.

At low redshifts ($z\lesssim 800-900$) the region up to $\pm 4$ to $\pm 10$
Doppler width seems to be quite important.
As explained in Sect.~\ref{sec:F_time}, there the correction is mainly because
of self-feedback, which is strongest where photons really are reabsorbed.
However, at practically all other redshifts one can clearly see that the
distant wings contribute significantly. Even within $\pm 10^3$ Doppler width
the deviations of the spectrum from the quasi-stationary solution are
important.
}

\subsection{Dependence of the escape probability on the shape of the
  emission and absorption profile}
\label{sec:P_shape}
\change{
As mentioned in the introduction, in the Sobolev approximation it is well known
that the result for the escape probability does not depend on the shape of the
Lyman $\alpha$ emission profile.
Looking at the derivation of expression \eqref{eq:Delta_n_CR_b} for $P_{\rm
  S}$ it is clear that in addition to the condition of quasi-stationarity one
needs $\phi_{\rm em}\equiv \phi_{\rm abs}\equiv \phi_{\rm sc}$, i.e. the
\changeF{equality} of the line emission, line absorption, and line scattering profile.
This conclusion is also reached in the case of no line scattering
(Sect.~\ref{sec:ncoh}) leading to $P_{\rm d}$ as given by Eq.~\eqref{app:Pd}.

However, if $\phi_{\rm em}\neq \phi_{\rm abs}$ then the situation is more
complicated.
Starting with Eq.~\eqref{app:Eq_Nnu_coh}, but allowing $\phi_{\rm em}\neq
\phi_{\rm abs}$, one can find the solution
\bsub
\label{app:kin_abs_em_Sol_phys_asym}
\beal
\label{app:kin_abs_em_Sol_phys_asym_a}
\Delta N^{\rm a}_{\nu}(z)&=[N_{\rm em}(z)-N^{\rm pl}_{\nu_{21}}(z)]\!\!\int_{z_{\rm s}}^z \!\!\Theta^{\rm a}(z')
\,
\partial_{z'}e^{-\tau_{\rm abs}(\nu, z', z)}\id z',
\\
\label{app:kin_abs_em_Sol_phys_asym_b}
\Theta^{\rm a}(z')
&=\frac{\tilde{N}_{\rm em}(z')
\times\frac{\phi_{\rm em}(z', \nu_{z'})}{\phi_{\rm abs}(z', \nu_{z'})}-\tilde{N}^{\rm pl}_x}
{\tilde{N}_{\rm em}(z)-\tilde{N}^{\rm pl}_{x_{21}}}
\end{align}
\esub
with $\nu_{z'}=\nu\,\frac{1+z'}{1+z}$, and $\tau_{\rm abs}(\nu, z', z)$ as
defined by Eq.~\eqref{app:tau_abs_all} \changeF{but replacing $\phi\rightarrow\phi_{\rm abs}$}.
With this, the mean occupation number in the Lyman $\alpha$ line can be
directly computed leading to
\bsub
\label{app:Nbar_kin_abs_em_Sol_asym}
\beal
\label{app:Nbar_kin_abs_em_Sol_asym_a}
\bar{n}^{\rm coh}(z)&=\bar{n}^{\rm pl}+\frac{c^2}{2\nu^2_{21}}
\int^\infty_0\varphi_{\rm abs}(\nu) \,\Delta N^{\rm a}_{\nu}(z)\id\nu
\nonumber
\\
&=\bar{n}^{\rm pl}+\Delta \bar{n}_{\rm em}(z)
\!\!\int^\infty_0 \!\!\varphi_{\rm abs}(\nu) \id\nu \!\!\int_{z_{\rm s}}^z \Theta^{\rm a}(z')\,
\partial_{z'} e^{-\tau_{\rm abs}(\nu, z', z)} \id z'
\nonumber
\\
&=\bar{n}^{\rm pl}+\Delta \bar{n}_{\rm em}(z)\,(1-P^{\rm a}_{\rm em})
\\[2mm]
\label{app:Nbar_kin_abs_em_Sol_asym_b}
P^{\rm a}_{\rm em}
&=
1+\int^\infty_0 \!\!\!\!\varphi_{\rm abs}(\nu)\id\nu \int^{z_{\rm s}}_z 
\!\!\Theta^{\rm a}(z')
\,
\partial_{z'}e^{-\tau_{\rm abs}(\nu, z', z)}\id z',
\end{align}
\esub
with the same definitions as in Eq.~\eqref{app:Nbar_kin_abs_em_Sol}. 
%
%
\changeF{Note that here one has to compute $\bar{n}$ using the absorption profile function.}
%

Now it is easy to show that for $\Theta^{\rm a}=1$, $\tau_{\rm
  abs}=\pdeath\tauS\,\chi$, and $\phi_{\rm em}\equiv \phi_{\rm abs}=\phi$, one
directly obtains the solution Eq.~\eqref{app:nb_coh}. Here the crucial point
is that it is possible to introduce the variable $\chi(\nu)=\int_0^\nu
\phi(\nu')\id \nu'$, so that there is no direct dependence on $\phi$.
\changeF{However, it is clear that already for $\phi_{\rm
  em}\neq \phi_{\rm abs}$, in general $\Theta^{\rm a}\neq 1$, implying that 
this coordinate transformation can only be  achieved
approximately, so that the result will depend on the shape of the
Lyman $\alpha$ line.}
Similarly, in the case $\phi_{\rm em}\equiv \phi_{\rm abs}$, but for
$\Theta^{\rm a}\neq 1$ and $\tau_{\rm abs}\neq f(z)\,\chi$, where $f(z)$ is
a function of redshift only, the result for the escape probability will depend
on $\phi$.
In that case, it will be important how large the deviations from
quasi-stationarity are in the range, where most of the contributions to $P$ originate. If for example the profile is a pure Doppler core, then any
photon that is emitted will at most travel, scatter and redshift over a
characteristic length $\Delta\nu/\nu\sim \text{few}\times 10^{-5}$ \changeF{before escaping}. This
corresponds to $\Delta z/z\sim 10^{-5}$, so that the properties of the medium
have not changed very much, and the correction should be $\Delta N_{\rm
  e}/N_{\rm e}\sim \text{few}\times 10^{-3}\%$.
However, if, as in the real problem, radiative transfer is occuring in the
distant Lorentz wings, \changeF{these} corrections will be important.
}

\begin{figure}
\centering \includegraphics[width=0.9\columnwidth] {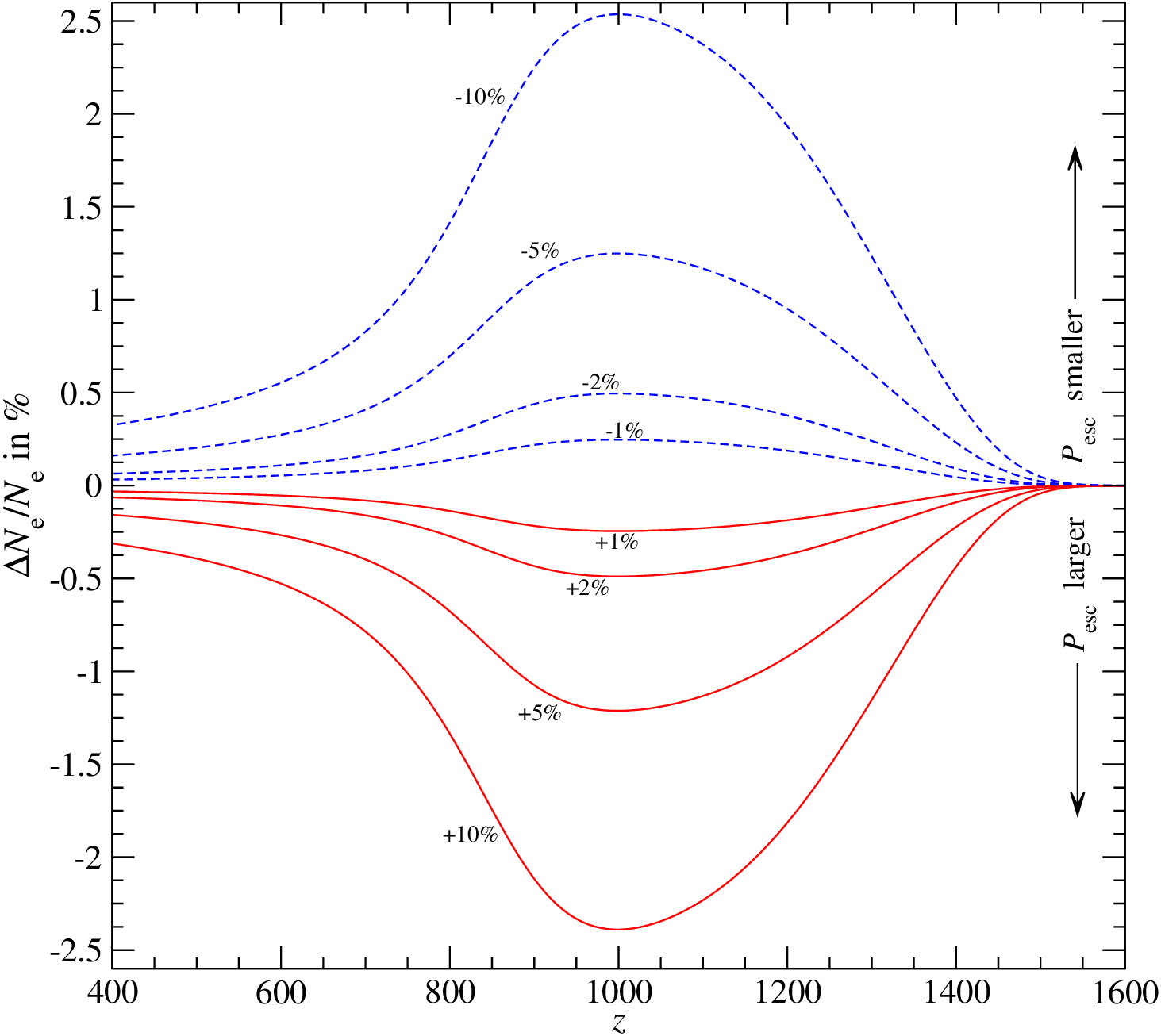}
\caption
{Relative change in the number density of free electrons when allowing for a
  constant relative change in the Sobolev escape probability of the
  Lyman $\alpha$ transition. At maximum $\Delta N_{\rm e}/N_{\rm e}\sim
  -\frac{1}{4}\Delta P/P$.}
\label{fig:DNe.Pesc}
\label{fig:DNe.DA_A}
\end{figure}
\section{Corrections to the ionization history}
\label{sec:changes_in_Xe}
With the results of the previous Section it is possible to estimate the
expected changes in the ionization history.
In Fig.~\ref{fig:DNe.Pesc} we show how a constant difference in the
Sobolev escape probability affects the ionization history.
We can see that the response is roughly proportional to the given $\Delta
P/P$.
Therefore we can use the curve for $\Delta P/P=1\%$ to estimate the changes in
the ionization history for the results given above.
Since all the corrections are small, one expects a small additional
correction, when computing the escape probability for this modified ionization
history.
For the purpose of this paper this approximation is sufficient.

Fig.~\ref{fig:DNe.Pesc} also shows that percent level corrections to the
escape probability do not affect the ionization history at $z\gtrsim 1600$,
while at $z\sim 1000$ one has $\Delta N_{\rm e}/N_{\rm e}\sim
-\frac{1}{4}\Delta P/P$.
Also one can see that at low redshifts, changes of the escape probability are
not propagating very much to the ionization history, resulting only in $\Delta
N_{\rm e}/N_{\rm e}\sim -0.054\,\Delta P/P$ at $z\sim 600$.
Note that the correction of $N_{\rm e}$ in both cases is much smaller than the
one of the escape probability. This is because the 2s-1s two-photon decay
channel already contributes slightly more to the effective recombination rate.

\begin{figure}
\centering
\includegraphics[width=0.9\columnwidth]{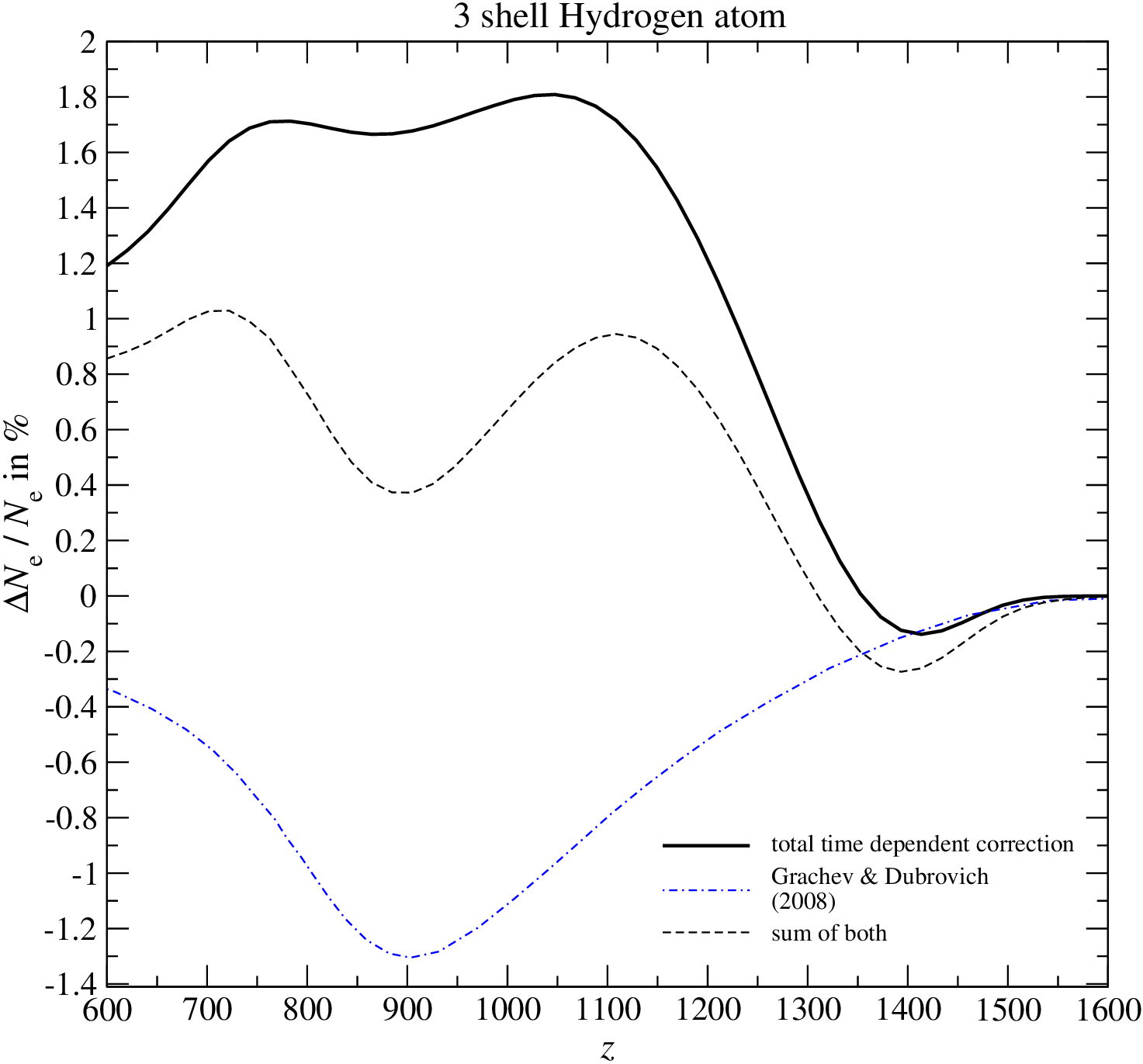}
\caption
{Estimated relative changes in the number density of free electrons when
  including various physical processes. The curves were obtained by simply
  multiplying the computed change in the escape probability for the 10 shell
  case as given in Fig.~\ref{fig:DP_P.time.nS} by the curve in
  Fig.~\ref{fig:DNe.Pesc} for $\Delta P/P=+1\%$ (thick solid line). 
We also show the result of \citet{Grachev2008} and the resulting sum of both
(dashed).}
\label{fig:DNe.predict}
\end{figure}
In Fig.~\ref{fig:DNe.predict} we give the correction of the number density of
free electrons as a function of redshift.
The time-dependent correction of the escape probability leads to a $\sim
1.6-1.8\%$ change of $\Delta N_{\rm e}/N_{\rm e}$ in the redshift range
$800\lesssim z \lesssim 1200$.
This change is practically twice as large as the effect due to line recoil
\citep{Grachev2008}, so that the sum of the time-dependent correction and the
recoil correction is still dominated by the former contribution, leaving
$\Delta N_{\rm e}/N_{\rm e}\sim 0.94\%$ at $z\sim 1100$.
This will be important for the computation of the CMB power spectra, where at
large $l$ in the case of $TT$ one expects a $\Delta C_l/C_l \sim 1\%$, and
about 2 times more for $EE$.

\section{Discussion and conclusions}
\label{sec:disc_con}

\subsection{Main results related to the cosmological ionization history and
  the CMB power spectra}
\label{sec:disc_con_0}
In this paper we investigated the validity of the Sobolev approximation for
the Lyman $\alpha$ escape probability during hydrogen recombination.
We separate absorption and emission of Lyman $\alpha$ photons from resonant
scattering events, including the fact that processes leading to full
redistribution of photons over the Voigt profile occur with much lower ($\sim
10^{-3}-10^{-4}$ times) probability than resonant scatterings.
We have shown that within the standard formulation the rapid changes in the
ionization degree \changeF{during} recombination lead to significant departures of the
photon distribution from the quasi-stationary solution.
We took these corrections into account analytically, assuming that the photon
redistribution process over frequency during a scattering is coherent in the
lab frame.

Although one does expect some additional modifications when accounting for
partial frequency redistribution, our computations show \citep{Chlubaprep}
that the additional correction will be dominated by the influence of
line recoil, that has been addressed in \citet{Grachev2008}.
However, the time-dependent correction that is considered here turns out to be
significantly larger, so that we focused on this only.
A more complete consideration of this problem is in preparation
\citep{Chlubaprep}.

Here we found that the time-dependent corrections to the effective
Lyman $\alpha$ escape probability result in a $\sim 1.6-1.8\%$ change of
$\Delta N_{\rm e}/N_{\rm e}$ in the redshift range $800\lesssim z \lesssim
1200$ (See Sect.~\ref{sec:Pesc_def_num_Voigt_time} and \ref{sec:changes_in_Xe}
for more detail).
These corrections are important for the Thomson visibility function and in
computations of CMB power spectra, where at large \changeF{($l \sim 1000-3000$)}
multipoles $l$ in the case of
$TT$ one expects modifications of the order of $\Delta C_l/C_l \sim 1\%$, and
about 2 times more for $EE$.
However, note that we also expect additional changes when formulating the
problem more rigorously in the two- or multi-photon approach (see discussion
in Sect.~\ref{sec:disc_con_1}).

The main reason for the corrections \changeF{discussed here} are (i) time-dependent changes in the
absorption optical depth, and (ii) changes in the net emission rate due to the
time-dependence of cosmological recombination.
The correction \changeF{due to case (i)} is especially important for contributions coming from the
distant wings of the Lyman $\alpha$ line, where emitted photons can travel,
scatter, and redshift over a very long time before getting reabsorbed. For the
3 shell hydrogen atom the associated correction to the escape probability is
$\Delta P/P\sim -1.2\%$ at $z\sim 1100$, while it results in an increase of
$\Delta P/P\sim +2.8\%$ at $z\sim 1440$ (for more details see
Fig.~\ref{fig:DP_P.time.P0}).

The correction related to the time dependence of the net emission rate is slightly larger, leading to a total decrease of the effective escape probability at
$z \lesssim 1400$, that for the 3 shell atom reaches $\Delta P/P\sim -5.8\%$
at $z\sim 1140$, while it results in an increase of $\Delta P/P\sim +3.7\%$ at
$z\sim 1510$ (for more details see Fig.~\ref{fig:DP_P.time}).
Here a significant contribution is due to departures of the very distant \changeF{($\sim 10^3-10^4$ Doppler widths)} wing
spectrum from the quasi-stationary solution (e.g. see
Sect.~\ref{sec:P_wings}).

\change{
  We also showed that in particular at low redshifts ($z\lesssim 800-900$)
  this correction, \changeF{owing to a self-feedback process}, 
  strongly depends on the number of shells that have been
  taken into account for the computation (see Fig.~\ref{fig:DP_P.time.nS}).
This is because the effective emission rate depends on the solution for the
populations \changeF{of} the excited levels, so that it is important to include at least
$4-5$ shells into the computations.
However, this aspect of the solution appears to be due to the incompleteness
in the formulation of the problem, so that at these redshift the conclusions
should change when using a two- or multi-photon description (see discussion in
Sect.~\ref{sec:disc_con_1}).
Also it is important to mention that for the corrections in the CMB power
spectra, this should not affect the results very much.
}

\subsection{Apparent problems with the standard formulation}
\label{sec:disc_con_1}
Our analysis shows that under the extreme physical conditions valid in the hot
Universe (extremely low plasma density in the presence of the intense CMB
radiation field), the standard formulation of the Lyman $\alpha$ transfer
problem leads to several apparently unphysical results.
First, we would like to point out that all our computations and estimates
clearly show how important (at the percent level accuracy) the distant wings of
the lines are for the value the escape probability or mean intensity
supporting the 2p level (e.g. see Sect~\ref{sec:distance} and
\ref{sec:P_wings}).
However, in the standard approach variations of the blackbody and also any
power-law variations in $\nu$ are usually neglected in the formulation of the
transfer problem and analytic computation, an approximation that is certainly
questionable when going to $|\Delta\nu/\nu|\gtrsim 1\%$, or $\sim 10^3$
Doppler width.
For example, as we mentioned in Sect.~\ref{sec:abs_term}, this approximation
leads to a small non-conservation of a blackbody spectrum at large distances
from the line center, an aspect that simply follows from exact application of
the detailed balance principle, \changeF{leading to a thermodynamic correction factor $f=\frac{\nu_{21}^2}{\nu^2}\,e^{h[\nu-\nu_{21}]/k\Tg}$. In a two-photon formulation of the problem this factor automatically appears \citep{ChlubaprepII}}.

Also, the emission of photons according to the standard Voigt profile in
principle allows the production of photons until arbitrarily large distances on
the blue side of the Lyman $\alpha$ resonance.
Without introducing some high frequency cut-off, in the cosmological
recombination problem these photons will lead to some unphysical self-feedback
at low redshift (e.g. see Sect.~\ref{sec:Pesc_def_num_Voigt_time_1}), which is
also present in our current solution, but at times that are not so important
for the CMB power spectra.
We expect that both problems can be resolved when using a two- or multi-photon
formulation, in which detailed balance is applied self-consistently, and where
the line profiles are naturally bound \citep[e.g. see][]{Chluba2008a} due to energy conservation.

Focusing on the Sobolev approximation (quasi-stationarity of the spectrum and
complete redistribution), several unphysical aspects also appear.
These are again due to the unique properties of our Universe, where there are
hardly any collision and the expansion rate is so low that the Sobolev optical
depth $\tauS$ reached values of $\sim 10^6-10^8$ during recombination.
As explained in Sect.~\ref{sec:Sobolev} this leads to the case that the
variations of the photon distribution that are important for the mean
intensity supporting the 2p level during cosmological recombination occur at
distances of $\sim 10^5-10^8$ Doppler widths from the resonance. This is far
beyond the Lyman $\beta$ line or even the ionization energy of the hydrogen
atom.
It is also possible to compute the present day Lyman $\alpha$ spectral
distortion in the time-dependent approach, using the solution~\eqref{app:kin_abs_em_Sol_CR_phys}, for which it was assumed that every
line scattering leads to a complete redistribution of photons over frequency.
We checked that in this case one would obtain a Lyman $\alpha$ line profile
that is very different from the one computed in the usual $\delta$ function
approximation \citep[e.g. see][]{Jose2006}.
One reason for this is that the effective frequency beyond which the photon
distribution is no longer affected by the Lyman $\alpha$ resonance is very far
on the red side of the line center (at $\xD\sim -10^3$ to $-10^4$, or
$|\Delta\nu/\nu|\sim 1\%-10\%$).
\change{
This aspect of our computations also suggests that in principle it should be
possible to constrain the type of redistribution that is at work during
hydrogen recombination by looking at the exact position and shape of the
residual Lyman $\alpha$ distortion in the CMB.
In all cases, no line scattering, complete redistribution, and partial
redistribution the Lyman $\alpha$ distortion will look different.

We conclude that for the conditions during cosmological recombination,
complete redistribution for a line scattering event in the standard
formulation is not an appropriate redistribution process, and leads to rather
unphysical results.
%
%
With the approximation of coherent scattering in the rest-frame, some of the
unphysical aspects of the solution disappear. However, as mentioned above, at
low redshifts we obtain a large feedback of Lyman $\alpha$ photons
initially released at high redshifts (e.g. see
Sect.~\ref{sec:Pesc_def_num_Voigt_time_1}).
These problems can be resolved using a two- or multi-photon formulation.
}

\subsection{Future prospects}
\label{sec:disc_con_2}
In spite of all the complications we expect that to lowest order one can take the
time-dependent correction during cosmological recombination into account using
the solution~\eqref{app:Nbar_kin_abs_em_Sol}, or in the full two-photon
description using a time-dependent solution in the {\it no-scattering
  approximation}.
Since all additional corrections will (also) be small, one can then compute
each other process more or less separately. 
This should also be possible for the case of helium recombination, but here
the reabsorption of photons by the small fraction of neutral hydrogen atoms
present at that time will be much more important \citep{Switzer2007I,
  Kholupenko2007, Jose2008}.

In order to include the final correction into the computations of the CMB
power spectra it will be \changeF{necessary} to develop a fast scheme for the evaluation
of the ionization history. For this purpose \citet{Fendt2008} recently
proposed a new approach called {\sc
  Rico}\footnote{http://cosmos.astro.uiuc.edu/rico/}, which uses
multi-dimensional polynomial regression to accurately represent the dependence
of the free electron fraction on redshift and the cosmological parameters.
Here one first has to produce a grid of models using a given full
recombination code, for which each run may take several hours or up to days.
However, the time-consuming part of the computation is restricted to the {\it
  training} of {\sc Rico}, while afterwards each call only takes a small
fraction of a second.
This approach should allow one to propagate all the corrections in the ionization
history that are included in the full recombination code to the CMB power
spectra, without using any \changeF{fudge-factors}, like in {\sc Recfast}
\citep{SeagerRecfast1999, Wong2008}.
In the future, we plan to provide an updated training set for {\sc Rico},
including the time-dependent corrections discussed here. This should also make
it easier for other groups to cross-validate our results.

\begin{appendix}

\section{Computational details}
\label{app:comp_details}

\subsection{Computations of $\phi(\nu)$}
The evaluation of the Voigt profile, Eq.\eqref{app:Voigt}, is usually rather
time-consuming.
However, convenient approximations for $\phi(\xD)$ can be given in the very
distant wings and also close to the center of the line.
For $|\xD|\leq 30$ we use the approximation based on the Dawson integral up to
sixth order as described in \citet[][Sect.~9.2, p.~279]{Mihalas1978}.
In the distant wings of the line ($|\xD|\geq 30$) we apply the Taylor expansion
\beal
\label{app:Voigt_appr_wing}
\phi_{\rm wings}
\!\approx\!\frac{a}{\pi x^2_{\rm D}}\!\left[1\!+\!\frac{3-2a^2}{2 x^2_{\rm D}}
+\frac{15-20a^2}{4 x^4_{\rm D}}+\frac{105(1-2a^2)}{8 x^6_{\rm D}}\right].
\end{align}
We checked that the Voigt function is represented with relative accuracy
better than $10^{-6}$ in the whole range of frequencies and redshifts.
Using Eq. \ref{app:Voigt_appr_wing}, on the red side of the resonance one can
approximate the integral $\chi=\int_{-\infty}^{\xD}\phi(x') \id x'$ by:
\beal
\label{eq:Int_Wings_appr}
\chi_{\rm wings}=-\frac{a}{\pi x_{\rm D}}\left[1+\frac{3-2a^2}{6 x^2_{\rm D}}
+\frac{3-4a^2}{4 x^4_{\rm D}}+\frac{15(1-2a^2)}{8 x^6_{\rm D}}\right].
\end{align}
as long as $x \lesssim -30$. Since $a\sim 10^{-4}-10^{-3}$, this shows that
in the distant wings only a very small fraction of photons is emitted.
Using the symmetry of the Voigt profile one finds $\chi(x)=1-\chi(-x)$, such
that Eq.~\eqref{eq:Int_Wings_appr} is also applicable for $x\gtrsim 30$.

\end{appendix}

\acknowledgement{JC wishes to thank Dimitrios Giannios, Stefan Hilbert and
  Stuart Sim for useful discussions.
  \changeF{The authors are also very grateful to E.~Switzer for carefully reading the manuscript and providing detailed comments.
Furthermore the authors would like} to thank Prof. Dubrovich, C.~Hirata,
E.E.~Kholupenko, J.A.~Rubino-Martin, and W.Y.~Wong for very
interesting discussions during the workshop 'The Physics of Cosmological
Recombination' held in Garching, July 2008.
}

\bibliographystyle{aa} 
\bibliography{Lit}

\end{document}